\documentclass[10pt]{wlscirep}
\usepackage[utf8]{inputenc}
\usepackage[T1]{fontenc}
\usepackage{bm}
\usepackage{amsmath}
\usepackage{comment}
\usepackage{listings}
\usepackage{xcolor}
\definecolor{codegray}{gray}{0.95}
\definecolor{green}{rgb}{0.07,0.88,0.58}
\definecolor{pink}{rgb}{0.88,0.22,0.68}
\usepackage[most]{tcolorbox}
\newcommand{\emphasis}[1]{{\textcolor{pink}{\textbf{#1}}}}
\usepackage{wrapfig}
\usepackage{hyperref}
\hypersetup{
    colorlinks=true,
    linkcolor=pink,
    citecolor=pink,
}
\usepackage{cleveref}

\title{Normative and neural network models of \\ multisensory integration}

\author[1, *]{Marcus Ghosh}
\author[2]{John S. Butler}
\affil[1]{Department of Life Sciences, Imperial College London, London, United Kingdom}
\affil[2]{School of Mathematics and Statistics, Technological University Dublin, Grangegorman Lower, Dublin, Ireland}

\affil[*]{e-mail: m.ghosh@imperial.ac.uk}

\begin{abstract}
Animals continuously detect scents, sights and sounds, which their brains combine into diverse multisensory streams. Multisensory integration accelerates learning, increases reaction speeds and helps to resolve sensory ambiguities. In studying this process, the field has developed a continuum of models; from concise mathematical equations, with a handful of variables, to artificial neural networks, with billions of parameters. In this review, we divide this spectrum into two classes: normative models start from a hypothesis about how multisensory signals should be integrated, while neural network models aim to emulate how circuits could integrate them. Together, these models demonstrate that observers can integrate cues in a statistically optimal fashion, and help to explain experimental findings, including the diversity of multisensory neuron response profiles and network architectures. In this review, we provide a didactic overview of multisensory models, complete with accessible equations and Python code, and compare their assumptions and explanatory power. We detail what has been learned from these models. Finally, we discuss how the ideas and insights from classical theoretical models can be reconciled with those from artificial neural networks, and how this convergence should shape future research.    
\end{abstract}

\begin{document}

\flushbottom
\maketitle

\thispagestyle{empty}

\newpage{}
\section*{Introduction}

Everything is multisensory. At an environmental level, natural habitats are rich with scents, sights and sounds; And, even in controlled laboratory settings, an animal performing a "unisensory" task will still detect other sensory signals (or the unnatural lack thereof). At the neural level, sensory representations are seemingly less segregated than classically thought. Multisensory processes are observable in primary cortical areas \cite{ghazanfar2006, driver2008,bizley2016}. And it is well established that there are anatomical connections between primary sensory cortices \cite{falchier2002, cappe2005} as well as feedback connections from higher multisensory areas back to "unisensory" areas \cite{driver2008}. These functional and anatomical results support the idea that \hyperref[glossary]{multisensory integration} is a ubiquitous and fundamental neural computation \cite{stein2008multisensory}.

However, multisensory integration is not any single computation. Instead, how senses are integrated likely depends on which pairs or combinations of modalities are being combined as well as which extracted features. For instance, in self-motion perception, vision primarily provides information about position and velocity, whereas the vestibular system provides information about velocity, acceleration, and jerk \cite{butler2015optimal}. And, audio-tactile  experimental data has shown multisensory interactions for frequency \cite{yau2009temporal} but not loudness \cite{yau2010separate} or duration \cite{butler2011common}.

Exploring multisensory integration is important as it would improve our knowledge of sensory processing, sensory-motor transformations and multi-region computations. Allowing us to develop better artificial multisensory systems \cite{baltruvsaitis2018, liang2022} and to begin to tackle the neurodevelopmental \cite{dionne2015} and neuropsychiatric conditions \cite{hornix2019} in which multisensory integration is thought to be altered and may even contribute to symptoms.

Understanding means building models. That is, if we want to figure out a complex process, we need to break it into simpler parts and to describe the relations between those parts \cite{beer2024, ghosh2025}. For experimentalists, models often take the form of "box and arrow" diagrams, supported by experimental data: activity in brain region A $\rightarrow$ behaviour. Theorists aim to characterise these links ($\rightarrow$), by developing models which explain how circuits can generate neural dynamics which drive appropriate behaviours. For theorists, models take numerous forms, from concise mathematical equations, with a handful of variables, to artificial neural networks, with billions of parameters. 

In studying multisensory integration, researchers have developed models along the width of this continuum. In this review, we divide this spectrum into two classes: normative models --- which start from a theory about how multisensory signals \textit{should} be integrated, and artificial neural networks which aim to emulate how circuits \textit{could} integrate them. This distinction is meaningful and helpful for three reasons. First, the \textit{motivation} for the two is often fundamentally different. Normative models emphasise a "top-down" scientific approach --- starting from a principle and trying to figure out if and how that principle could be implemented by neural circuits. Conversely, neural network models support a "bottom-up" approach --- beginning from circuits and building towards an understanding of the computations they perform. Second, normative and neural network models can differ in their \textit{interpretability}. Most normative models use only a handful of parameters, so are highly interpretable --- in a two parameter case, for instance, one can plot a models fit as function of both. By contrast, many neural network models use thousands to millions of parameters, which makes them more challenging to interpret. Third, the two are often \textit{validated} on different criteria. In general, normative models are validated based on their ability to fit experimental data, while neural network models are judged on their ability to perform multisensory tasks, akin to those used to test experimental subjects. That said, individual models can blur this distinction. For instance, some mechanistic models test the role of specific computations, such as divisive normalisation, using neural network simulations, and there are emerging methods which integrate normative and neural network elements. 

Beyond improving our knowledge of multisensory integration, by helping to explain empirical data, models are valuable as they can be used to: test how well an analysis method works before applying it experimentally \cite{zhu2026}, guide experimental design \cite{jones2016}, and to generate new experimentally testable hypotheses. 

In this review, we provide a didactic introduction to normative and neural network models of multisensory integration. We conclude by making the case that the most promising direction for the field will be to focus on research which integrates the two, to develop interpretable models that are capable of capturing the complexities of naturalistic multisensory behaviours and multi-region neural activity.   

\newpage
\begin{figure}[h!]
    \centering
    \includegraphics[width=0.95\linewidth]{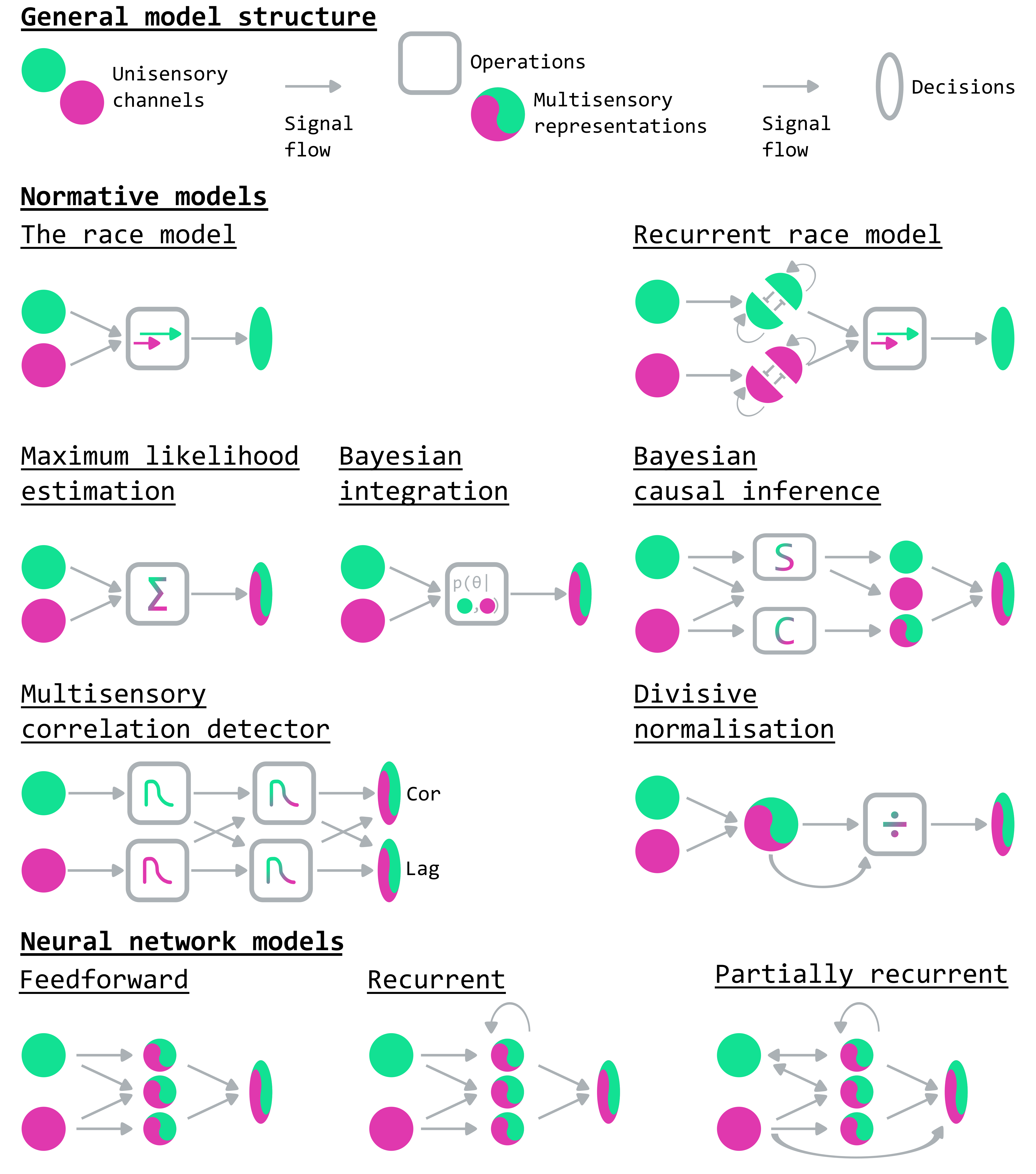}
    \caption{Normative and neural network models of multisensory integration. All models are shown with the same general structure; unisensory signals (green and pink circles) contribute to, one or more, normative operations (squares) or multisensory representations (green and pink circles), the outputs of which inform the model's decisions (vertical ellipse). All models are included here to facilitate model-to-model comparisons. Each is shown again and explained in more detail later on.}
    \label{fig:ModelZoo}
\end{figure}

\newpage
\section*{Normative models}
\subsection*{Additive Model}

Normative models start from a hypothesis about how multisensory signals \textit{should} be integrated, and are then evaluated for their ability to explain experimental data. For instance, Meredith and Stein (1986) used an additive model (Equation \ref{ADDITIVE}) to describe the firing rates (FR) of neurons in the cat Superior Colliculus \cite{meredith1986visual}:
\begin{equation}
\label{ADDITIVE}
    FR_{Add} = FR_A+FR_{V}
\end{equation} 
    \begin{equation}
\label{Percent}
    Percent=\left(\frac{FR_{AV}-FR_{Add}}{FR_{Add}}\right)*100
\end{equation}  
where $FR_A$, $FR_V$, and $FR_{AV}$ denote the neuron's firing rates in response to auditory, visual, and audiovisual stimuli, respectively, and $FR_{\text{Add}}$ is the firing rate predicted by the additive model. The percentage measure (Equation \ref{Percent}) quantifies the extent to which the observed audiovisual response differs from the additive prediction. 

\begin{figure}[h]
    \centering
\includegraphics[width=0.6\linewidth]{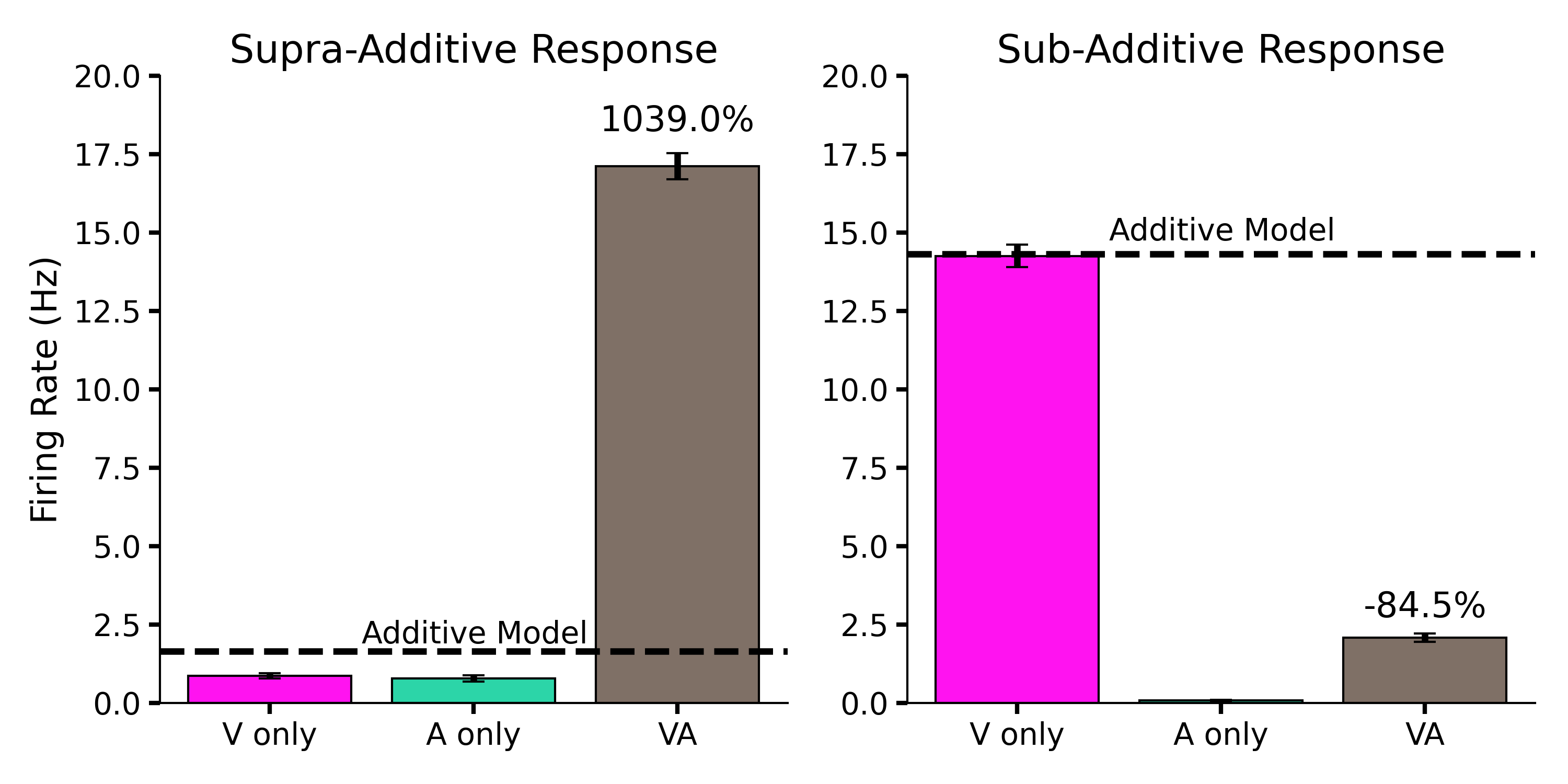}
    \caption{Simulated firing rate of neurons in the cat Superior Colliculus to visual-only (pink), auditory-only (green) and audio-visual (brown) stimuli, and an additive model following Equation \ref{ADDITIVE} (dotted line). Based on results from \cite{meredith1986visual}. Left: Supra-Additive. Right: Sub-Additive }
    \label{fig:FiringRateAdd}
\end{figure}

Although very few neurons' firing rate responses were exactly predicted by this additive model, it introduced a framework for classifying the firing rate of multisensory neurons as \hyperref[glossary]{supra-additive} if the multisensory response is more than the sum of the unisensory responses (Figure \ref{fig:FiringRateAdd} left) or \hyperref[glossary]{sub-additive} if the multisensory response is less than the sum of the unisensory firing rates (Figure \ref{fig:FiringRateAdd} right).  

In Python, we could compare a neuron's unisensory and multisensory firing rates, using the following function (Listing \ref{code:add}).

\begin{lstlisting}[
    caption={Python code for the additive model.},
    label={code:add}
    ]
import numpy as np 

def firing_rate_add(FR_A,FR_V,FR_AV):
    """
    Compare a neuron's audio-visual firing rate to the sum of its unisensory responses.

    Arguments: 
        FR_A is a numpy array of firing rates, from multiple trials for a neuron, in response to an auditory stimuli.
        FR_V is a numpy array of firing rates, from multiple trials for a neuron, in response to a visual stimuli.
        FR_AV is Numpy vector with the firing rates, from multiple trials for a neuron, in response to multisensory audio-visual stimuli. 
    Returns:
        Percent difference between the mean audio-visual firing rate and the
        mean of the summed auditory + visual rates. Negative values indicate
        sub-additive responses, positive values supra-additive.
    """
     
    FR_Add = FR_A + FR_V 
    
    Percent=(np.mean(FR_AV)-np.mean(FR_Add))/np.mean(FR_Add)*100

    return Percent
\end{lstlisting}

Identifying neurons with supra- or sub-additive responses implies that there is an interaction between sensory streams (a multisensory process) but does not propose an underlying mechanism. That being said, a recent study in mouse frontal cortex showed that this area combined auditory and visual evidence in an additive fashion which mirrored the animal's behaviour \cite{coen2023}. And, a modelling study simulating auditory and visual reaction times found that an additive model produced the optimal speed-up on multisensory trials \cite{brady2025modelling}.  

\subsection*{Race Models}
\begin{figure}[ht]
    \centering
\includegraphics[scale=0.8]{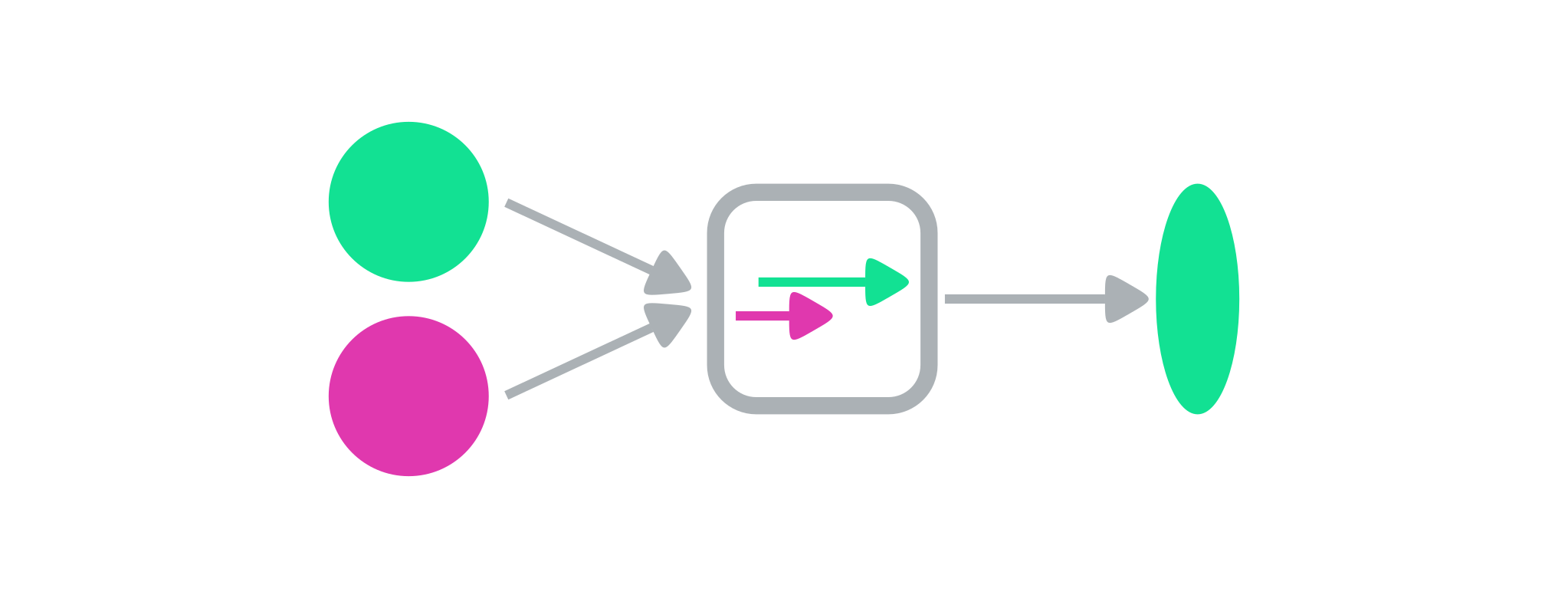}
    \caption{In the race model, each modality is processed independently (green and pink circles and arrows), and observers base their decision (vertical ellipse) on whichever sense is fastest (here the green sense).}
    \label{fig:Race_model}
\end{figure}

The race model has been applied to investigate multisensory integration in reaction time data \cite{raab1962, miller1982, otto2017, roberts2025}. Conceptually, each sense "races" to produce an independent decision, and whichever is fastest wins the race (Figure \ref{fig:Race_model}). Technically, it defines a bound of the probabilistic summation based on two or more unisensory reaction time distributions \cite{diederich2004bimodal}; which accounts for the fact that even without sensory integration, choosing the fastest decision from multiple sensory channels will lead to faster responses than choosing from a single modality.

In its standard formulation, the race model bound is expressed as \cite{miller1982}:
\begin{equation}
\label{RM_Simple}
    F_{AV}(t)\leq F_A(t)+F_{V}(t),
\end{equation}  
where $F_A(t)$ and $F_V(t)$ denote the fraction of trials which finish before time $t$ (the unimodal cumulative reaction time distributions) and $F_{AV}$ is the cumulative multisensory reaction time distribution. Observing faster reaction times than this bound (Equation \ref{RM_Simple}) therefore implies that multisensory responses cannot be explained by probability summation alone, and instead reflect co-activation between sensory channels \cite{miller1982}.

A stricter bound can be derived under the assumption of stochastic independence between sensory channels, meaning that trial-to-trial fluctuations in auditory and visual processing occur independently of one another. In practice this means incorporating an additional term to Equation \ref{RM_Simple}:
\begin{equation}
\label{RM_Indep}
    F_{AV}(t)\leq F_A(t)+F_{V}(t)\textcolor{pink}{-F_A(t)F_{V}(t)},
\end{equation}  
which corresponds to the cumulative distribution of the minimum of two independent processes.

\begin{figure}[h!]
    \centering
\includegraphics[width=0.9\linewidth]{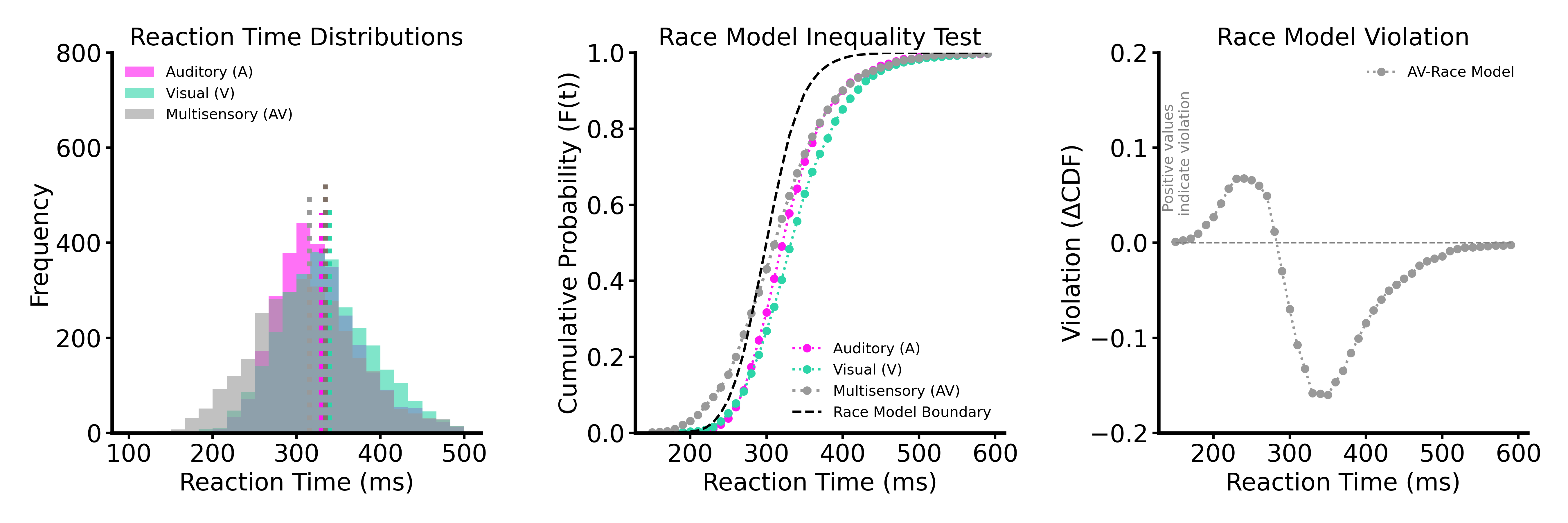}
    \caption{Reaction time analyses across auditory (A), visual (V), and audio‑visual (AV) conditions.
Left: Histograms show the distribution of reaction times for auditory, visual, and audio‑visual conditions.
Center: Cumulative distribution functions (CDFs) illustrate the race‑model inequality test. The dashed line indicates the race‑model boundary, with the multisensory CDF exceeding the boundary at shorter latencies.
Right: Race‑model violation plot ($\Delta$CDF) showing positive deviations (violations) for the multisensory condition relative to the race‑model prediction and corresponding linear comparison. Positive values represent moments where multisensory responses are faster than allowed by statistical facilitation. Simulated results based on \cite{brandwein2013development}.}
    \label{fig:ReactionTimeRace}
\end{figure}

Figure \ref{fig:ReactionTimeRace} illustrate the steps in calculating the Race Model Bound. The unisensory audio (pink) and visual (green) reaction time distributions (Figure \ref{fig:ReactionTimeRace} left) are converted to cumulative distribution functions (Figure \ref{fig:ReactionTimeRace} middle), which represent the proportion of trials as a function of reaction time. The unisensory cumulative reaction time distributions are then used to generate the bound using Equation \ref{RM_Simple} or \ref{RM_Indep}. This bound is then subtracted from the multisensory data to calculate the race model violation (Figure \ref{fig:ReactionTimeRace} right), positive values represent moments where multisensory responses are faster than expected by statistical facilitation. In Python, we could write Listing \ref{code:race}.

\noindent 
\begin{minipage}{\linewidth}
\begin{lstlisting}[
    caption={Python code to calculate the Race Model Boundary and its violations.},
    label={code:race}]
import numpy as np

def get_cdf(data, bins):
    # Compute the cumulative distribution function (CDF).
    return [np.mean(data <= t) for t in bins]

def race_model_bound(RT_a,RT_v,RT_av):
    """ 
    Arguments:
        RT_a: an array of reaction times (ms) to an auditory stimuli.
        RT_v: an array of reaction times (ms) to a visual stimuli.
        RT_av: an array of reaction times (ms) to an auditory-visual stimuli.
    Returns: 
        race_model_boundary: will be an array of the Race Model Boundary over time. 
        violation: will be the difference between the Race Model Boundary and the cummulative audio-visual distribution over time. 
    """
    # Define reaction-time bins from 150 ms to 590 ms in 10 ms increments.
    # These bins are used to evaluate the cumulative distribution functions (CDFs).
    
    time_bins = np.arange(150, 600, 10)
   
    # Calculate CDFs for the auditory, visual, and audiovisual conditions.  
    F_a = np.array(get_cdf(RT_a, time_bins))
    F_v = np.array(get_cdf(RT_v, time_bins))
    F_av = np.array(get_cdf(RT_av, time_bins))
    
    race_model_boundary = F_a + F_v - F_a * F_v  # Race Model Bound (Equation 4)
    
    violation = F_av - race_model_boundary
    
    return race_model_boundary, violation
\end{lstlisting}
\end{minipage}

Otto and colleagues \cite{otto2013principles,otto2017, roberts2024semantic} extended the race model to account for the congruency of stimuli and the repetition/switch of trials, which can distort the interpretation of multisensory integration \cite{SHAW2020122}. And, Colonius \& Diederich (2017) proposed a version of the race model to characterise neuronal spiking \cite{colonius2017measuring}. In this context, the inequality provides a bridge between systems-level neurophysiology and behaviour: exceeding the race model bound implies that the combined sensory inputs produce a greater-than-expected impact on the decision process, consistent with integrative neural mechanisms. 

\subsection*{Probabilistic Models}
\begin{figure}[h]
    \centering
\includegraphics[scale=0.8]{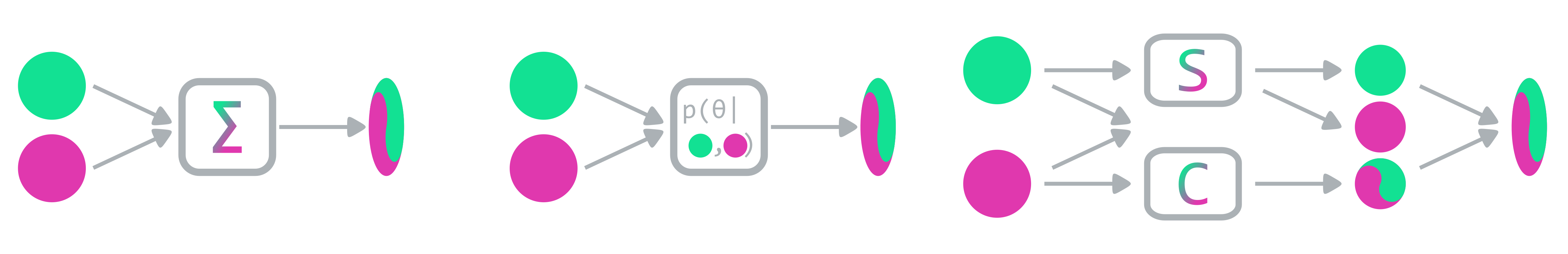}
    \caption{\textbf{Probabilistic models}. In maximum-likelihood estimation (left) each modality (green and pink circles) is weighted and linearly combined ($\sum$) to form a multisensory estimate (multicolour vertical ellipse). In Bayesian inference (middle) sensory inputs are additionally combined with prior probabilities ($\theta$). Causal Bayesian inference (right) adds a prior which considers if the sensory signals share a common source and should be combined ($C$) or independent sources and should be separated ($S$).}
    \label{fig:Bayesian_models}
\end{figure}

The introduction of maximum-likelihood estimation (MLE)  provided a predictive framework for multisensory modelling \cite{yuille1993bayesian, Anastasio2000BayesSC,alais2004ventriloquist}. Ernst and Banks (2002) demonstrated that adults integrate multisensory cues in a statistically \hyperref[glossary]{optimal} manner \cite{ernst2002} consistent with MLE. 
Consider a visual estimate $\hat{S}_{\mathrm{Vis}}$ and a vestibular estimate $\hat{S}_{\mathrm{Vest}}$ \cite{butler2010bayesian,butler2011role}. MLE (Figure \ref{fig:Bayesian_models} left) predicts that the optimal multisensory estimate $\hat{S}_{\mathrm{VisVest}}$ is a weighted sum of these unisensory estimates (Equation \ref{MLE}):
\begin{equation}
\label{MLE}
    \hat{S}_{VisVest}=w_{Vis}\hat{S}_{Vis}+w_{Vest}\hat{S}_{Vest},
\end{equation}
where each weight is proportional to the reliability (inverse variance) of its cue, $w_i = \frac{1/\sigma_i^2}{\sum_j 1/\sigma_j^2}$, assuming the sources are conditionally independent. Intuitively, sensory cues with lower variance are given greater weight because they provide more reliable estimates of the external world.

\begin{figure}[h!]
    \centering
\includegraphics[width=0.9\linewidth]{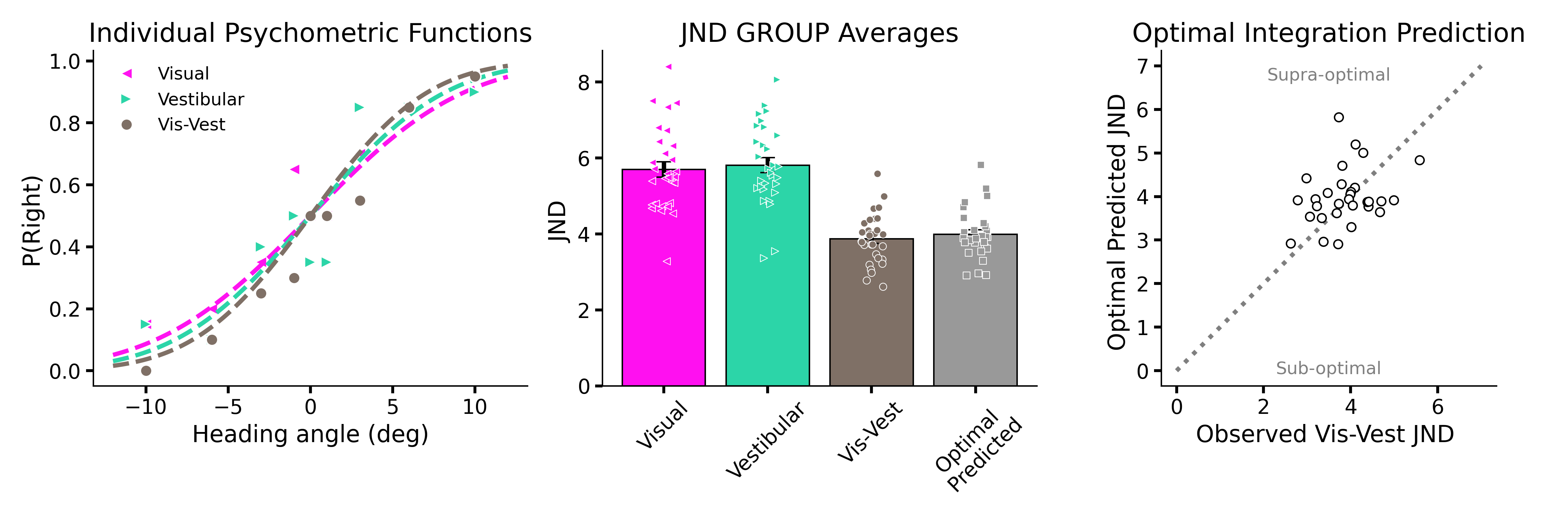}
 \caption{\textbf{Fitting and testing the MLE model of multisensory integration.} (\textbf{Left}) Individual psychometric functions for an example participant, showing the proportion of "rightward" responses as a function of heading angle for unisensory visual (pink) and vestibular (green) cues, and for the combined visual-vestibular (Vis-Vest) condition (brown). Dashed lines show cumulative Gaussian fits, from which the point of subjective equality (mean) and just-noticeable difference (JND; related to $\sigma$) are extracted for each condition. (\textbf{Middle}) Group-averaged JND for the visual, vestibular, and Vis-Vest conditions, alongside the JND predicted by optimal cue integration (Equation \ref{eq:MLE_var}) given the unisensory variances. Error bars represent SEM. The empirically observed Vis-Vest JND closely matches the optimal prediction, consistent with statistically optimal cue combination. (\textbf{Right}) Predicted versus observed JND for individual participants (open circles), with the unity line denoting perfect agreement between model and data. Points falling close to the line indicate that individual-level multisensory benefit is well predicted by MLE. 
 Simulations based on \cite{butler2011role}.}
    \label{fig:Bayes}
\end{figure}

In practice, these parameters are estimated by fitting a psychometric function to unisensory and multisensory response data (Figure \ref{fig:Bayes} Left). The mean, also known as the \hyperref[glossary]{point of subjective equality} and the variance, $\sigma^2$, which is proportional to the \hyperref[glossary]{just noticeable difference} (Figure \ref{fig:Bayes} Middle) are extracted for each condition (Listing \ref{code:MLE}). The fitted values are used to observe and predict the unisensory weights and the optimal reduction in variance for the multisensory condition (Figure \ref{fig:Bayes} Right), given by Equation \ref{eq:MLE_var}:
\begin{equation}
   \sigma_{VisVest}^2=\frac{\sigma_{Vis}^2\sigma_{Vest}^2}{\sigma_{Vis}^2+\sigma_{Vest}^2}.
   \label{eq:MLE_var}
\end{equation}

Notably, this MLE-predicted weighting has also been observed at the neuronal level \cite{Anastasio2000BayesSC,gu2008neural}, lending biological support to the model. A related approach extends the MLE framework to predict the optimal accumulation of evidence over time during multisensory integration \cite{Drugowitsch2014}. While MLE bases its estimate on the available sensory evidence, Bayesian integration also incorporates what the observer expects before observing that evidence (Figure \ref{fig:Bayesian_models} middle). These expectations are shaped by previous experience and bias perception towards stimulus values that are more commonly observed.

Although MLE and Bayesian integration successfully predict many multisensory phenomena, they both assume that cues share a common source and therefore cannot account for how integration is affected by spatial and temporal disparities. Such disparities influence whether cues are perceived as arising from a single source or from two distinct sources, with distinct behavioural \cite{Wallace2004Unifying} and neurophysiological \cite{gu2008neural} signatures. \hyperref[glossary]{Causal inference} models (Figure \ref{fig:Bayesian_models} right) address this by extending the Bayesian framework with a prior over whether signals originate from a common source ($C=1$) or from independent sources ($C=2$) \cite{Kording2007, sato2007bayesian, macneilage2007bayesian}. These models handle both cue combination and cue conflict, with the posterior probability of a common cause given by Bayes’ rule (Equation \ref{eq:Bayes}):
\begin{equation}
\label{eq:Bayes}
    \Pr(C | x_{V},x_{A})=\frac{ \Pr(x_{V},x_{A}|C)\Pr(C)}{\Pr(x_{V},x_{A})}.
\end{equation}
More recently, these causal inference processes have also been modelled using artificial neural networks \cite{terada2024chaotic,rideaux2021}.

\noindent 
\begin{minipage}{\linewidth}
\begin{lstlisting}[
    caption={Python code to calculate the Maximum Likelihood just noticeable difference for a Visual-Vestibular heading task.},
    label={code:MLE}]
def predict_optimal_combined_jnd(heading_angles, vis_prop_right, vest_prop_right,
                      visvest_prop_right):
    """
    Compute the MLE-predicted multisensory JND for a visual-vestibular
    heading task.

    Arguments:
        heading_angles: heading angles tested (deg)
        visual_prop_right: proportion of rightward responses at heading angles for the visual condition
        vest_prop_right: proportion of rightward responses at heading angles for the vestibular condition
        visvest_prop_right: proportion of rightward responses at headings angle visual-vestibular condition
`
        fit_psychometric_function fits a cumulative Gaussian and returns the
        point of subjective equality (PSE) and the just noticeable difference (JND).
        
    Returns:
        predicted_jnd: the predicted optimal multisensory just noticeable difference
        visvest_jnd: the observed visual-vestibular just noticeable difference
    """
    vis_pse, vis_jnd= fit_psychometric_function(
        heading_angles, vis_prop_right)
    vest_pse, vest_jnd = fit_psychometric_function(
        heading_angles, vest_prop_right)
    visvest_pse, visvest_jnd = fit_psychometric_function(
        heading_angles, visvest_prop_right)

    var_vis, var_vest = vis_jnd**2, vest_jnd**2
    predicted_jnd = np.sqrt(var_vis * var_vest / (var_vis + var_vest))
    
    return predicted_jnd, visvest_jnd
    
\end{lstlisting}
\end{minipage}

\subsection*{Mechanistic models}
Mechanistic models aim to explain experimental observations by invoking common neuronal level computations \cite{FROUDISTWALSH2026}.

\begin{figure}[h]
    \centering
\includegraphics[scale=0.8]{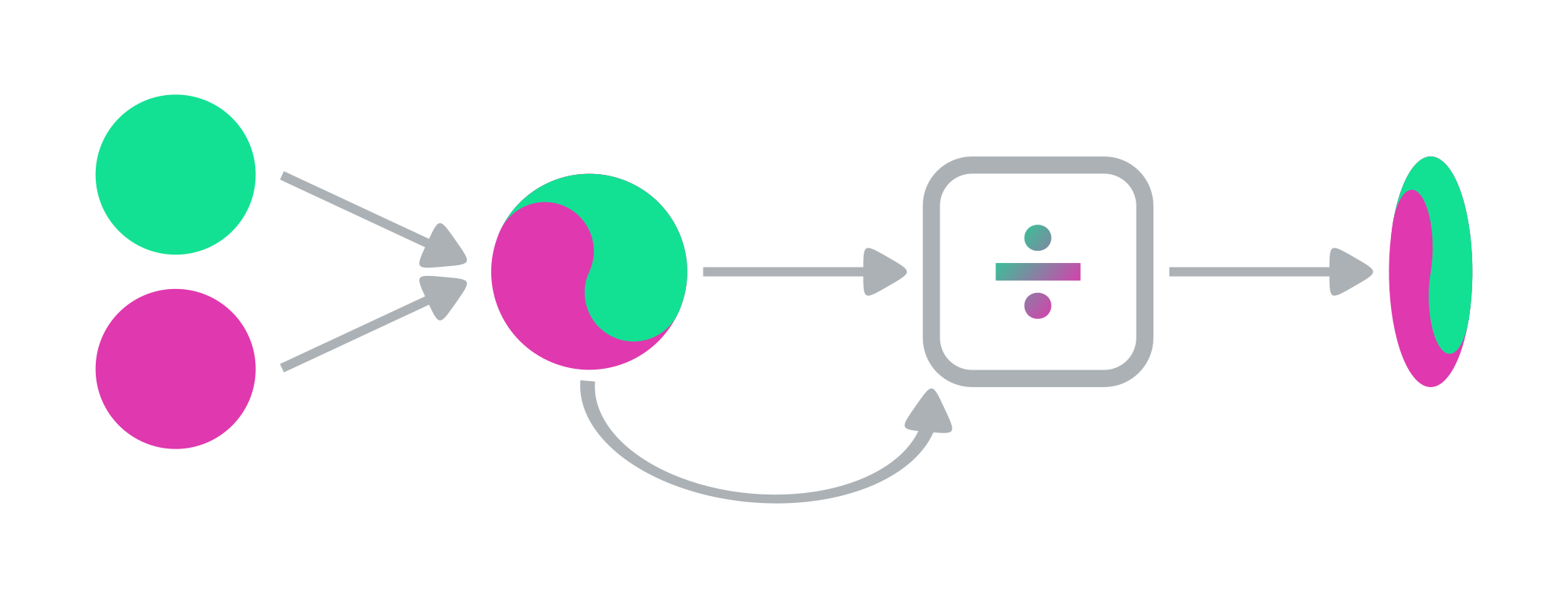}
    \caption{In divisive normalization unisensory signals (green and pink circles) are combined by a population of multisensory neurons (multicolour circle) whose activities are normalized relative to each other ($\div$) before contributing to decisions (multicolour vertical ellipse).}
    \label{fig:Divisive_normalisation}
\end{figure}

Divisive normalization has been proposed as a mechanism underlying the response profiles of multisensory neurons \cite{beck2011,Ohshiro2011}. According to this model, the activity of each multisensory neuron is divided by the combined activity of a multisensory population (Figure \ref{fig:Divisive_normalisation}). Ohshiro et al., (2011) showed that divisive normalization can account for classic empirical principles of multisensory integration, including inverse effectiveness as well as spatial enhancement and suppression \cite{Ohshiro2011}. Their two-layer model reproduced response patterns observed in neurons from the Superior Colliculus, ventral intraparietal area, and dorsal medial superior temporal area.

\begin{figure}[h]
    \centering
\includegraphics[scale=0.8]{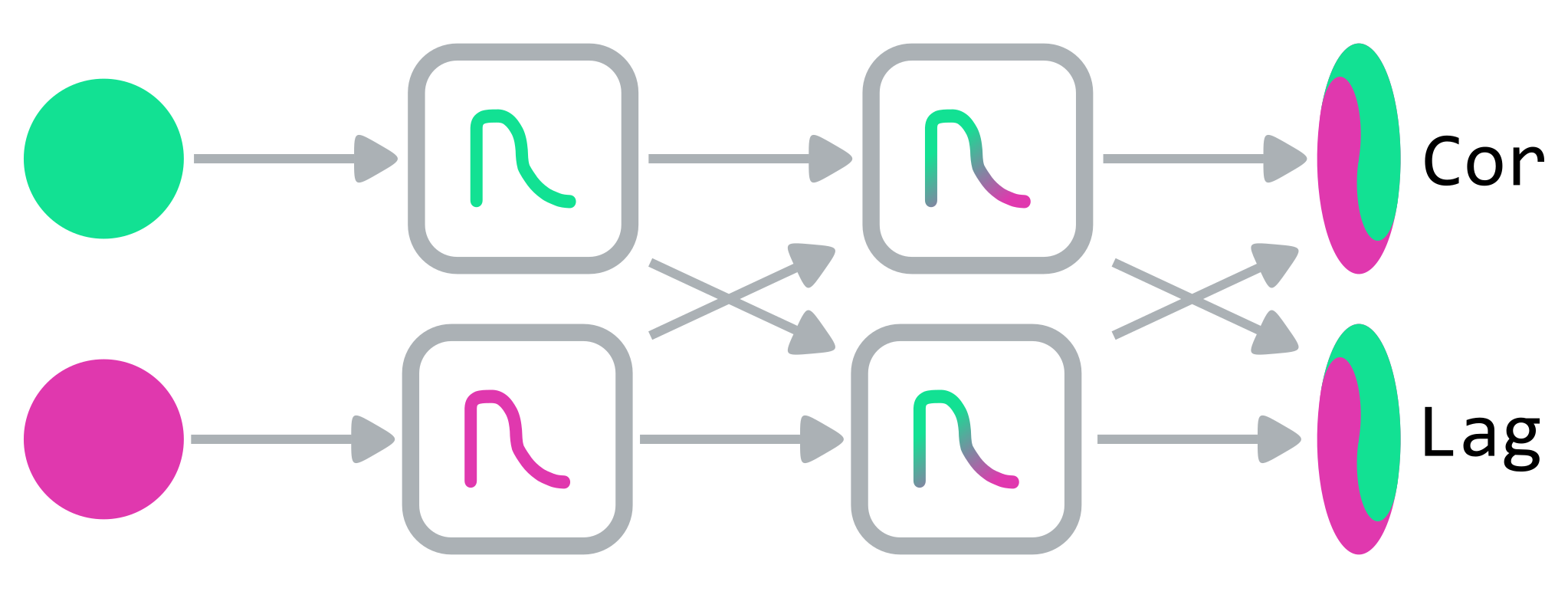}
    \caption{In the multisensory correlation detector, unisensory signals (green and pink circles) are processed by banks of temporal filters (squares). These outputs can then be used to compute the correlation and temporal lag between senses (multicolour vertical ellipses).}
    \label{fig:MCD}
\end{figure}

Another class of mechanistic models focuses on the temporal correlation between sensory signals as a basis for determining whether they should be integrated \cite{Parise2016, pesnot2022, parise2025}. Inspired by the Hassenstein–Reichardt motion detector \cite{HassensteinReichardt1956}, Parise \& Ernst (2016) proposed a multisensory correlation detector \cite{Parise2016}, in which modality-specific temporal filtering and multiplicative interactions detect cross-modal correlation, synchrony, and temporal lag (Figure \ref{fig:MCD}). This model reproduced a broad range of human psychophysical findings such as Bayesian-optimal cue integration while accounting for its breakdown under temporal conflict. 

A final class of mechanistic models use neural network simulations, in which the parameters (for instance, the strength of neuron-to-neuron connections) are set by hand; either based on experimental data, or by systematically exploring a range of values. These models allow for the consequences of different connection strengths or integration rules to be examined in response to simulated stimuli.  

Magosso et al., 2008 and Ursino et al., 2009 modelled multisensory integration in the Superior Colliculus using a neural network architecture in which two unimodal areas and a multimodal area are connected via both feedforward (uni-to-multisensory) and feedback (multi-to-unisensory) connections \cite{magosso2008theoretical,ursino2009}. Given a fixed set of parameters, these models exhibited \hyperref[glossary]{supra and sub-additivity}, \hyperref[glossary]{inverse effectiveness}, \hyperref[glossary]{cross-modal enhancement} and \hyperref[glossary]{cross-modal suppression}; demonstrating that these classical multisensory phenomena can all arise from a relatively simple network model.   

\begin{figure}[h]
    \centering
\includegraphics[scale=0.8]{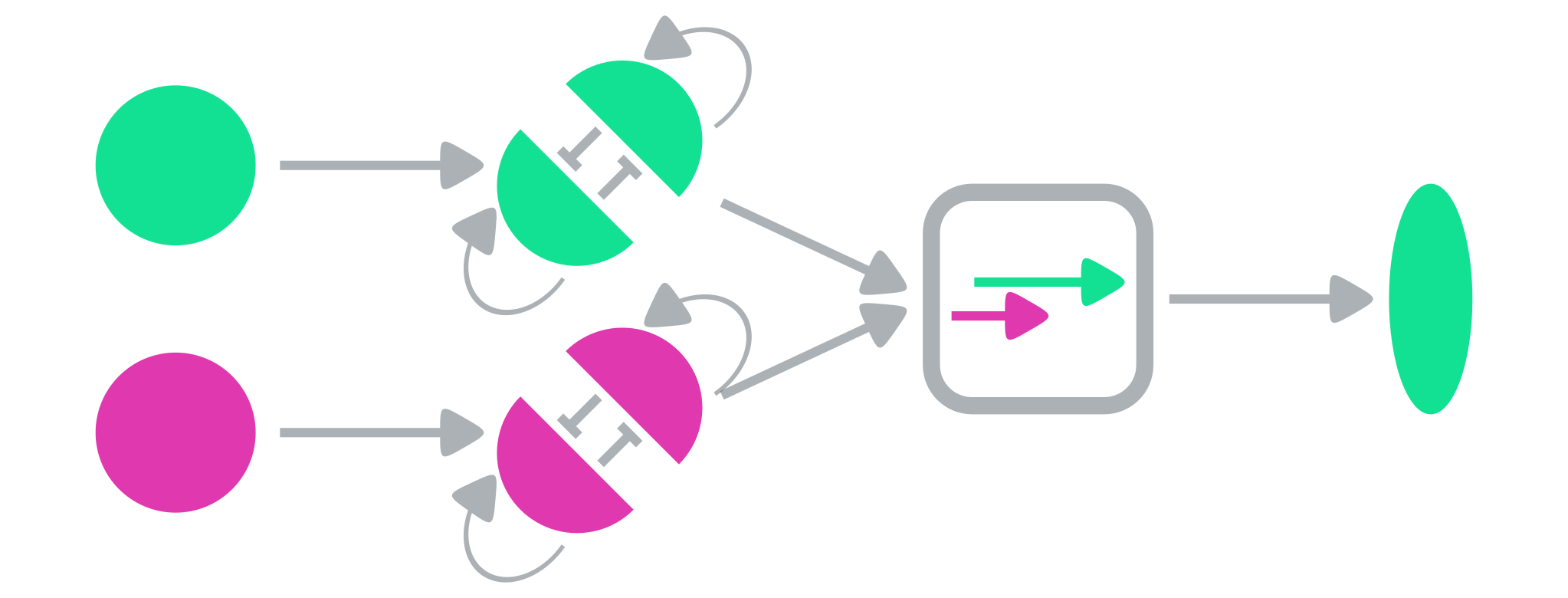}
    \caption{In the recurrent race model, each unisensory channel (green and pink circles) is processed by multiple populations of neurons (semi circles) --- with excitatory intra-population (grey arrows) and inhibitory inter-population connections (grey bars). The final decision (vertical ellipse) is based on whichever population reaches a decision first (here, one of the green ones).}
    \label{fig:Recurrent_race_model}
\end{figure}

Similarly, Brady \& Butler (2025) used recurrent mean-field models in which differential equations describe the temporal evolution of population firing rates (Figure \ref{fig:Recurrent_race_model}) \cite{brady2025modelling}. They showed that alternative architectures—including winner-take-all, linear integration, and history-dependent mechanisms—can reproduce multisensory response benefits while making distinct mechanistic assumptions. Like other mechanistic models, this approach links circuit-level architectures to behavioural signatures, but it additionally provides a framework for examining how the temporal dynamics of neural populations and alternative integration mechanisms may generate similar behavioural outcomes.

Overall, normative models of multisensory integration help to explain many experimentally observed phenomena. The main advantage of these models is that they are highly interpretable, as they make explicit assumptions and use few parameters. However, their main drawback, is that there is a limit to how well they can fit experimental data, and the level of task complexity which they can tackle. 

\section*{Neural network models}
Brains transform their sensory inputs by passing messages between groups of interconnected neurons. Analogously, artificial neural networks compute by passing signals through layers of interconnected units. Using techniques from machine learning \cite{bishop2024}, artificial networks can be trained to perform complex multisensory tasks, akin to those used to test experimental (human or animal) subjects. As these models are \textit{fully observable} their learned connectivity, neural activity and behaviour can then be explored to see if they recapitulate experimentally observed phenomena. Moreover, as these networks are \textit{fully malleable} they can be used to conduct \textit{in silico} experiments beyond what is possible \textit{in vivo}. In this way, neural network models can be used to test if existing experimental tools provide meaningful insights \cite{jonas2017}, to benchmark new methods \cite{fakhar2022} and to develop novel hypotheses. 

In this section, we review how networks with different architectures (i.e. patterns of neuron-to-neuron connectivity) have been used in this way, highlighting the insights from these studies.     

\subsection*{Feedforward networks}
\begin{figure}[h]
    \centering
\includegraphics[scale=0.8]{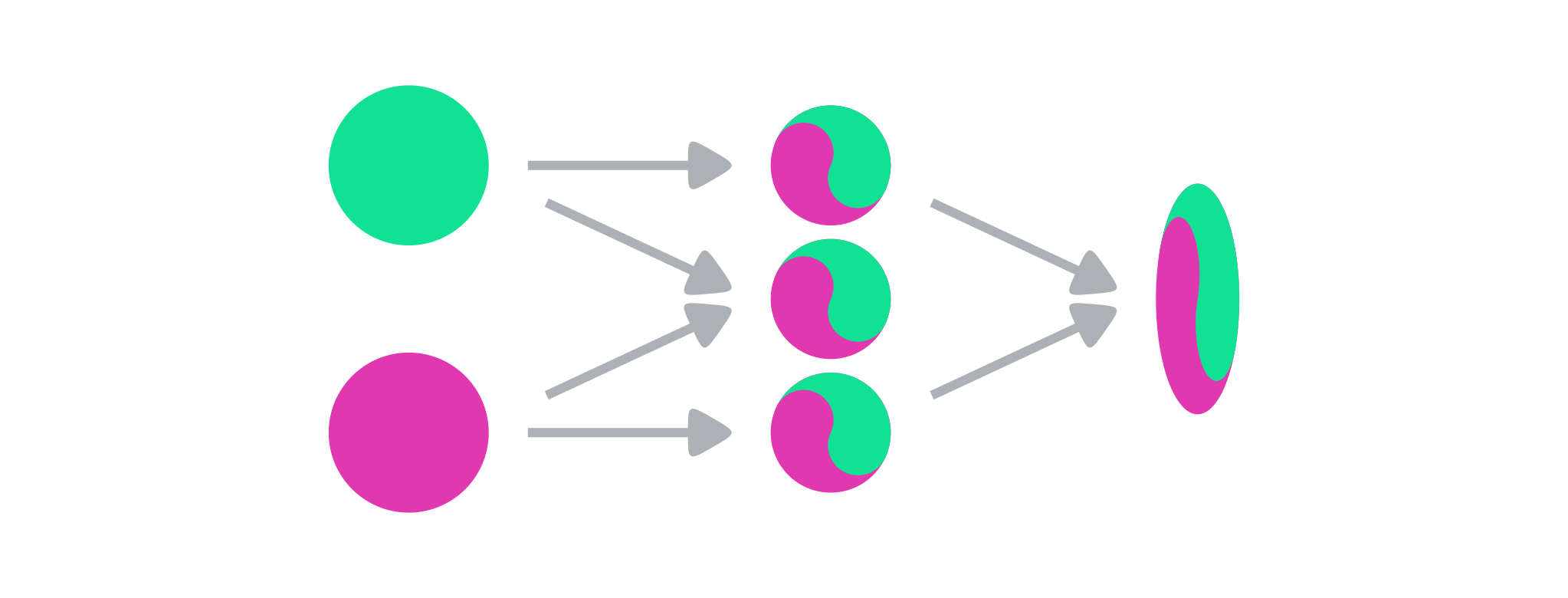}
    \caption{In feedforward neural networks, unisensory signals (green and pink circles) pass forward (grey arrows) to multisensory neurons (multicolour circles) and then decision outputs (multicolour vertical ellipse).}
    \label{fig:FF_model}
\end{figure}

In one of the simplest architectures, signals simply pass forward from sensory inputs to \hyperref[glossary]{hidden neurons} to decision outputs (Figure \ref{fig:FF_model}). 

We can model a feedforward network, with three populations of neurons (input - $i$, hidden - $h$ and output - $o$), mathematically as (Equation \ref{eq:ff}): 

\begin{equation} \label{eq:ff}
\hat{y} = g(W_{ho}f(W_{ih}x))
\end{equation}

Where $x$ is a vector representing the network's multisensory inputs, $W_{ih}$ and $W_{ho}$ are weight matrices which specify the strength of the connections between neurons ($W_{from-to}$), $f$ and $g$ denote \hyperref[glossary]{activation functions} which set how each neuron behaves, and $\hat{y}$ is a vector which represents the networks output. Note that we have omitted bias terms for clarity. In PyTorch \cite{paszke2019}, a package for building neural networks in Python, we could write (Listing \ref{code:ff}): 

\begin{lstlisting}[
    caption={PyTorch code for a feedforward neural network},
    label={code:ff}
    ] 
import torch
import torch.nn.functional as F

def feedforward(x, W_ih, W_ho, f=F.relu, g=F.softmax): 
    """
    Calculates a forward pass (input-to-output transformation) for a feedforward neural network. 
    Note: the shape of each argument and return is denoted in []. 
    Arguments: 
        x: a vector of inputs [inputs].
        W_ih: a weight matrix [hidden units, inputs].
        W_ho: a weight matrix [outputs, hidden units].
        f: an activation function for the hidden units.
        g: an activation function for the output units.
    Returns: 
        output: a vector of activations [outputs].
    """
    hidden = f(W_ih @ x)
    output = g(W_ho @ hidden, dim=0)
    return output
\end{lstlisting} 

By adjusting a neural network's parameters (in this case, the connection strengths in $W_{ih}$ and $W_{ho}$), we can train networks to transform their inputs ($x$) into specific outputs ($\hat{y}$) \cite{bishop2024}. Consequently, we can train networks to perform multisensory tasks. For instance, to choose left or right in response to directional stimuli such as random dot motion and lateralised audio signals.

By training feedforward networks on multisensory tasks, researchers have observed a diverse range of multisensory phenomena. At a behavioural level these networks can can optimally integrate the signals from multiple sensory channels \cite{Ghosh2024, farahmandi2024} and perform causal inference \cite{rideaux2021}. At a neuron level, researchers have identified \hyperref[glossary]{congruent and opposite neurons} \cite{rideaux2021}, neurons with sub, additive and supra-additive responses \cite{Ghosh2024, farahmandi2024}, and those which exhibit inverse effectiveness and cross-modal suppression \cite{farahmandi2024}. Together, these results demonstrate that several hallmark properties of multisensory integration can emerge in task-trained feedforward networks, without the need for some of the machinery proposed by normative models. For instance, divisive normalisation is typically thought to involve lateral or feedback connections and an explicit normalization pool \cite{beck2011, Ohshiro2011} and predictive multisensory processing would likely require top-down, feedback connections \cite{talsma2015}. This is not to say that these connection types are not present in multisensory circuits, but rather that feedforward networks are sufficient for these specific multisensory behaviours and neuron properties.    

Moving beyond observation, researchers have used virtual ablations to test the necessity of multisensory neurons with different properties for particular tasks. Specifically, Rideaux et al., 2021 demonstrated a role for both congruent and opposite neurons in causal inference \cite{rideaux2021}, and Ghosh et al., 2024 demonstrated the relative importance, in terms of task performance, of additive/supra-additive neurons in settings with dense/sparse multisensory signals \cite{Ghosh2024}. 

Overall, these results demonstrate that surprisingly many multisensory phenomena can be recapitulated by purely feedforward networks, and highlight the advantages of working with these observable and malleable models. However, as these models lack any memory or state, they are not well suited to tasks with naturalistic, time-varying multisensory signals nor modelling the complexities of neural dynamics \cite{chandrasekaran2017, senkowski2024, bolhasani2026}.

\subsection*{Recurrent networks}
\begin{figure}[h]
    \centering
\includegraphics[scale=0.8]{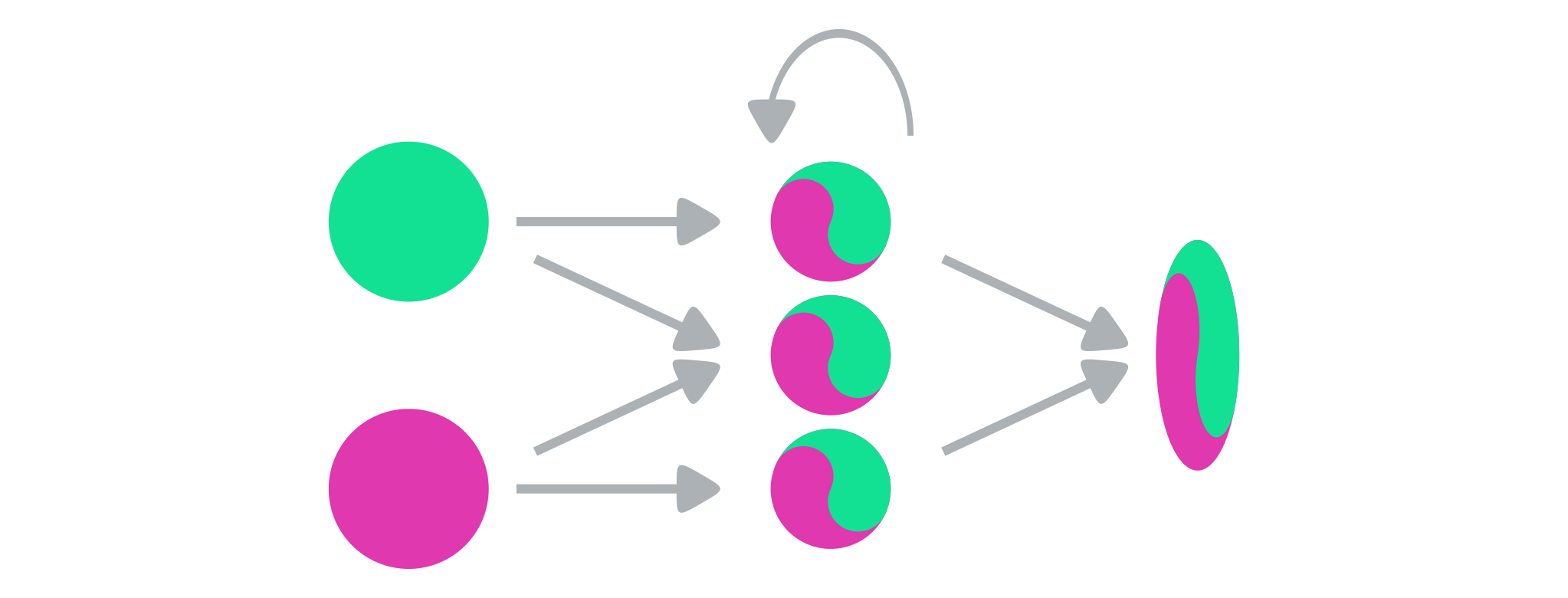}
    \caption{In recurrent neural networks, multisensory neurons (multicolour circles) combine incoming sensory signals (green and pink circles via grey left to right arrows) with the network's current state (grey right to left arrow). These multisensory activations shape the network's outputs (multicolour vertical ellipse).}
    \label{fig:RNN_model}
\end{figure}

The standard recurrent neural network (RNN) \cite{elman1990} adds an additional set of connections between the neurons in the hidden layer ($W_{hh}$), Figure \ref{fig:RNN_model}. This allows these neurons to combine incoming sensory signals ($x_t$) with the network's current internal state ($h_{t-1}$); the latter of which can act as a memory of the network's past inputs and hidden states.

Mathematically, we can introduce a single term to our feedforward equation (\ref{eq:ff}) to model an RNN (Equation \ref{eq:RNN}):

\begin{equation} \label{eq:RNN}
\hat{y_t} = g(W_{ho}f(W_{ih}x_t + \textcolor{pink}{W_{hh}h_{t-1}}))
\end{equation} 

In Python, we could modify our feedforward code as such (Listing \ref{code:rnn}): 
\begin{lstlisting}[
    caption={PyTorch code for a recurrent neural network},
    label={code:rnn}
    ]
import torch
import torch.nn.functional as F

def recurrent(x, h_tm1, W_ih, W_hh, W_ho, f=F.relu, g=F.softmax):
    """
    Calculates a single forward pass (input-and-state-to-output transformation) for a recurrent neural network. To process a sequence of inputs over time, this function would need to be called repeatedly; using inputs from the current time step (x_t) and hidden layer activations from the previous time step (h_tm1). 
    Note: the shape of each argument and return is denoted in [].
    Arguments: 
        x: a vector of inputs [inputs].
        h_tm1: a vector of hidden unit activations from the previous time step (time minus one) [hidden units].
        W_ih: a weight matrix [hidden units, inputs].
        W_hh: a weight matrix [hidden units, hidden units].
        W_ho: a weight matrix [outputs, hidden units].
        f: an activation function for the hidden units.
        g: an activation function for the output units.
    Returns: 
        hidden: a vector of activations [hidden units].
        output: a vector of activations [outputs].
    """
    hidden = f(W_ih @ x + W_hh @ h_tm1)
    output = g(W_ho @ hidden, dim=0)
    return hidden, output
\end{lstlisting}
     
At a behavioural level, Yamashita et al., 2013 demonstrated that an RNN with fixed, excitatory connections between nearby multisensory neurons and inhibitory connections between distant ones, can appropriately separate or integrate multisensory stimuli; depending on whether they come from the same or different sources \cite{yamashita2013}. However, a key advantage of the RNN model, is that it can learn to perform multisensory tasks with complex, time-varying signals. For example, Anil et al., 2025 developed a multisensory predator-prey task, in which the observer (acting as the predator) must estimate their prey's overall direction of motion, but the prey only emit sparse bursts of signals, as if they were darting between hiding places \cite{anil2025}. Notably, this task is already too complex for normative Bayesian models! That is, one can write down the solution, but solving it is computationally intractable \cite{ma2023, anil2025}. By contrast, feedforward networks can estimate the prey's motion at every time-step, and accumulating these estimates over time leads to reasonable performance. But, by leveraging their intrinsic memory to detect the prey's bursts, RNNs can perform significantly better \cite{anil2025}. This highlights the fact that studying naturalistic tasks, may quickly demand neural network rather than normative models.  

At a neural level, Wierda et al., 2025 showed that the multisensory units in an RNN learn heterogenous tuning patterns (e.g. to sensory or output variables) \cite{wierda2025}. And Terada \& Toyoizumi (2024) demonstrated that multisensory RNNs with chaotic neural dynamics, induced by applying a strong gain parameter to the hidden-to-hidden weight matrix ($W_{hh}$), can learn to output sequences of estimates whose distributions match Bayesian priors and posteriors \cite{terada2024chaotic}. 

Building on the standard RNN architecture and the complexities of real multisensory circuits, several studies have built models with additional connection pathways. Ursino et al (2009) added recurrent connections between the neurons in their model's auditory ($W_{aa}$) and visual areas ($W_{vv}$) and explored how within- and cross-modal suppression depend on the balance between the relative strength of their model's uni and multimodal recurrence \cite{ursino2009}. Similarly, both Magosso et al (2008) and Tugsbayer et al (2024) explored the role of feedback connections (from multisensory to unisensory areas, $W_{hi}$) in models with multiple recurrent areas \cite{ursino2009, tugsbayar2024}. Expanding on these ideas, Ghosh and Goodman (2026) systematically explored 128 multisensory circuits (termed partially recurrent neural networks) with different combinations of recurrent, feedback and skip connections \cite{ghosh2026}. They found that different architectures behave significantly differently (in terms of their task performance, learning speed and robustness to various perturbations), and explained these differences by demonstrating that different architectures learn solutions with distinct stimulus sensitivities and memory dynamics \cite{ghosh2026}.

\begin{figure}[h]
    \centering
\includegraphics[scale=0.8]{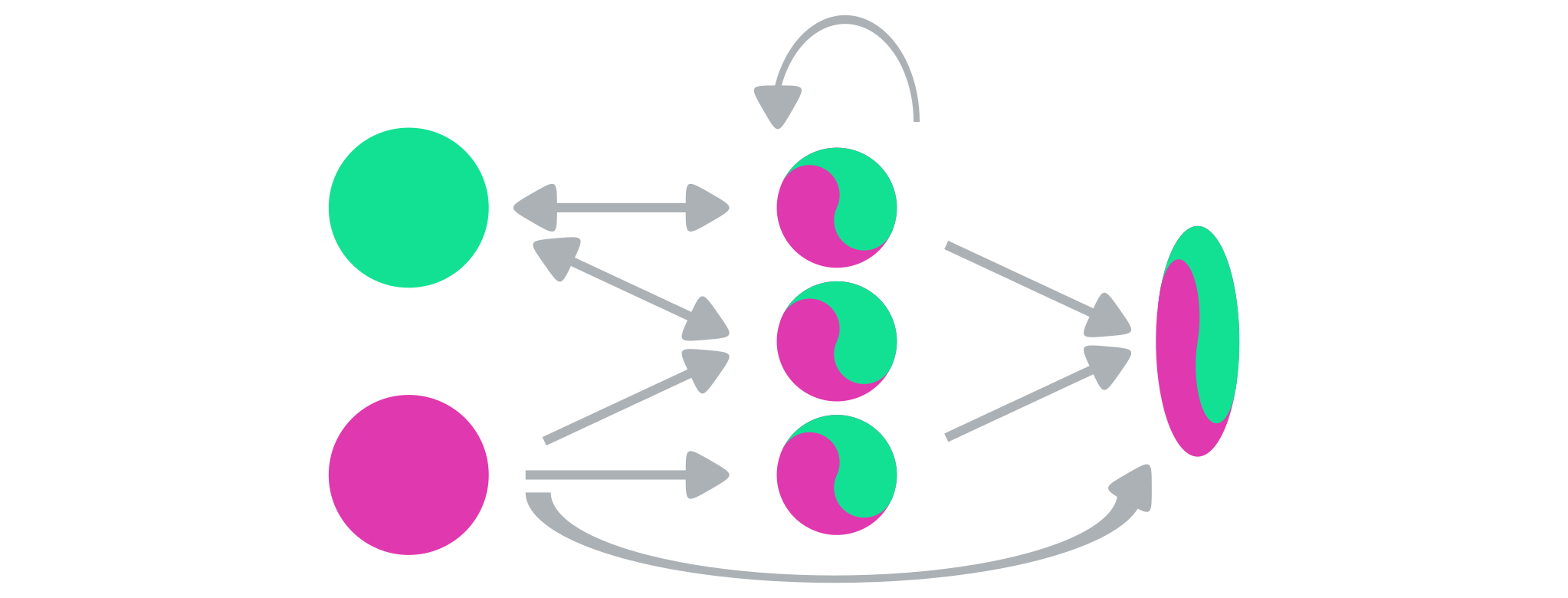}
    \caption{In partially recurrent neural networks \cite{ghosh2026}, signals propagate between multiple areas through different combinations of pathways. Here for example, signals pass forward from unisensory inputs (green and pink circles) to a set of multisensory neurons (multicoloured circles) and then decision outputs (multicoloured ellipse), but also: backwards from the hidden neurons to the green "unisensory" neurons, laterally between multisensory neurons and directly from the purple unisensory neurons to the decision outputs, skipping the hidden layer. Other partially recurrent neural networks would include different combinations of pathways, hence distinct patterns of signal flow.}
    \label{fig:pRNNs}
\end{figure}

\subsection*{Alternative architectures}
Advances in multimodal machine learning offer new architectures, concepts and tools for addressing open neuroscientific questions \cite{liang2022}. For instance, to explore whether modalities should be fused early or late \cite{bizley2016}, one can build, train and compare networks with different points of multisensory convergence. In this direction, early fusion has been shown to boost performance \cite{gadzicki2020} and robustness to noise \cite{barnum2020} in multimodal convolutional neural networks. And using multimodal transformers \cite{vaswani2017}, Nagrani et al. (2021) demonstrated that early, but restricted cross-modal communication can be both performant and energy efficient \cite{Nagrani2021}. 

\section*{Integrating normative and neural network models}
Above, we introduced many models of multisensory integration, and highlighted the insights which the field has gained from them. Now, looking ahead, we will consider \textit{what is the most we could hope to learn from normative and neural network models of multisensory integration}? 

By fitting \textbf{normative models} to experimental data, one can: see how well a given model fits the data, compare the fits between different models to determine which best explains the data, and interpret fit parameters to garner insights into the integration process. A stellar example of this approach comes from Coen et al., 2023 who taught mice to perform an audio-visual localisation task \cite{coen2023}. After training, their mice’s behaviour and neural activity were well accounted for by a relatively simple additive model, with limited evidence that more complex multisensory interaction terms improved explanatory power. 

The key strength of normative models is that they often use only a handful of parameters, so are highly interpretable. However, their main limitation is a phenomenon termed model degeneracy \cite{edelman2001, marder2011}. That is, one can imagine numerous normative models of multisensory integration and in practice, many models may explain the same data equally well. One solution to this problem, following Occam's razor, is to prioritise simpler models. For example, as basic neuron models with purely feedforward inputs and sigmoidal \cite{magosso2008theoretical, farahmandi2024} or leaky-integrate-and-fire \cite{Ghosh2024} input-output functions, can exhibit inverse effectiveness and generate sub- to supra-additive states, mechanisms such as divisive normalisation \cite{Ohshiro2011} or modulatory "feedback" \cite{oess2020} may be unnecessary for these multisensory phenomena. Another solution, is to design experiments which specifically distinguish between candidate models. For instance, Ghosh et al., 2024 introduced a family of nonlinear multisensory tasks, which facilitate a direct comparison of additive and nonadditive integration \cite{Ghosh2024}; in the most extreme setting, any observer performing significantly above chance must be using a nonadditive strategy. Looking ahead, the field could even develop generative adversarial collaborations --- where theories are explicitly tested head-to-head \cite{blohm2024, peters2025}.

By developing \textbf{neural network models} of multisensory integration one can tackle questions at each of Marr's three levels \cite{Marr1982, lengyel2024, pillow2024}. At the computational level, we can study what types of networks can implement which behaviours or even create simulated environments to explore the conditions in which multisensory integration becomes advantageous. At the algorithmic level, we can study the behavioural and neural algorithms which these models use --- using methods from neuroscience and machine learning interpretability; such as decoding \cite{glaser2020} or neuron-level ablations \cite{fakhar2022}. And at the implementation level, we can consider how feasibly an algorithm could be implemented by a network of neurons. For instance, most multisensory studies focus on a combination of two senses, but generating scaling laws for the number of senses, could help us to distinguish how plausible different models are. 

The main advantages of neural network models is their ability to learn complex psychophysical or naturalistic tasks, and to generate neuron or circuit-level predictions. One, often cited, limitation is their lack of interpretability or "black box" nature. However, this need not be the case. Recently, Ji-An et al., 2025 demonstrated that tiny recurrent neural networks (with 1-4 hidden neurons) can be trained to model trial-by-trial behavioural data and outperform classical cognitive models; while remaining small enough to be interpreted using methods from dynamical systems theory \cite{ji2025}. Similarly, Ghosh et al., 2024 showed that networks with two hidden neurons could be trained to perform multisensory psychophysical tasks \cite{Ghosh2024} and Brady \& Butler, 2025 simulated recurrent multisensory networks using only a handful of parameters \cite{brady2025modelling}. Even if more complex models are necessary, then they provide a foundation for developing and testing the tools which we will require to understand biological networks \cite{jonas2017}. A deeper challenge is that networks with different training data, architectures and hyperparameters (or even just random seeds) may learn solutions with different behaviours and neural representations. While this presents the opportunity to study how these components shape network function \cite{richards2019, doerig2023}, it also highlights the importance of reporting results from many neural network models across different random seeds and hyperparameter sets \cite{patterson2023, crilly2025}. Rather than reporting, \textit{our neural network model recapitulated a given behaviour}, we should aim to report \textit{how often and under what conditions our model} did so.   

Looking ahead, there are emerging methods which integrate normative and neural network models, to overcome their individual limitations. Eckstein et al., 2026 \cite{eckstein2026} combined neural networks with normative components, and used these hybrid models to model human reward learning. And Castro et al., 2025 leveraged a novel search algorithm --- in which a large language model oversees an evolutionary process \cite{romera2024}, to automatically discover normative models which accurately describe human and animal behaviour in a psychophysical task \cite{castro2025}. These exciting approaches suggest that the most promising future direction for the field may be to integrate the two, to develop models capable of capturing the complexities of naturalistic behaviour, while remaining interpretable. 

\section*{Conclusions} 
Ultimately, normative and neural network models offer complementary paths to understanding multisensory integration. Normative models clarify what computations an observer should perform under particular assumptions about uncertainty and causal structure. Neural network models illustrate how such computations could emerge from networks with particular architectures, objectives and learning rules. The challenge going forward is to build models that are not just accurate, but generate testable hypotheses: whose failures distinguish between theories, whose internal representations generate testable neural predictions, and whose assumptions can be experimentally manipulated. With this in mind, the most promising direction is likely not to choose between normative and neural network models, but to leverage and integrate both. 

\newpage{}
\begin{tcolorbox}[colback=pink!5!white, colframe=pink, title=Glossary]
\phantomsection
\label{glossary}
\begin{itemize}
\item \emphasis{Activation function} --- a transformation used to calculate the output of a neuron in an artificial neural network.  
\item \emphasis{Congruent neurons} --- multisensory neurons whose tuning preferences across modalities are aligned. E.g. a neuron preferring rightward visual and vestibular motion.
\item \emphasis{Crossmodal enhancement} --- the multisensory response exceeds the best unimodal response. 
\item \emphasis{Crossmodal suppression} --- the multisensory response is lower than the best unimodal response. 
\item \emphasis{Hidden neurons} --- the neurons in an artificial neural network which are neither input nor output units.
\item  \emphasis{Inverse effectiveness} --- a tendency for weaker unimodal stimuli to yield larger multisensory responses, this is generally associated with two weak unisensory responses. 
\item \emphasis{Just-Noticeable Difference} --- the smallest difference between two stimuli (e.g. in intensity, direction, or speed) that can be reliably discriminated by an observer, typically taken as a measure of perceptual (or neural) sensitivity/precision.
\item \emphasis{Multisensory Causal Inference} --- the brain's (implicit or explicit) inference of whether two or more sensory signals arise from a common underlying cause and should therefore be integrated, or from separate causes and should be treated (segregated) independently.
\item \emphasis{Multisensory Integration} --- the process by which the nervous system combines information from two or more sensory modalities (or multiple cues within a modality) into a single, typically more precise and reliable, perceptual estimate.
\item \emphasis{Multisensory Optimal Integration} --- a combined (multisensory) estimate that achieves the lowest possible variance (or JND) attainable by weighting each unisensory cue in inverse proportion to its own variance, as prescribed by maximum-likelihood/Bayesian cue combination.
\item \emphasis{Opposite neurons} --- multisensory neurons whose tuning preferences across modalities are anit-aligned. E.g. preferring rightward motion in one modality but leftward in another. 
\item \emphasis{Point of Subjective Equality} --- the stimulus value at which an observer perceives two stimuli (or a stimulus and a reference) as equal, typically the 50\% point on a psychometric function, used as a measure of perceptual bias.
\item \emphasis{Redundant Signal Effect (RSE)} --- the behavioural speeding (faster reaction times) and/or accuracy improvement observed when two or more redundant, task-relevant sensory signals are presented together compared to either signal alone.
\item \emphasis{Sub-additive} --- classification of a neuron's multisensory firing rate, the response to the multisensory stimuli is less than the sum of its unimodal responses.
\item \emphasis{Supra-additive} --- classification of a neuron's multisensory firing rate, the response to the multisensory stimuli is more than the sum of its unimodal responses.
\end{itemize}

\end{tcolorbox}

\newpage{}
\bibliography{references.bib}

@article{Wallace2004Unifying,
  author    = {Wallace, Mark T. and Roberson, Gwendolyn E. and Hairston, W. David and Stein, Barry E. and Vaughan, James W. and Schirillo, James A.},
  title     = {Unifying multisensory signals across time and space},
  journal   = {Experimental Brain Research},
  year      = {2004},
  volume    = {158},
  number    = {2},
  pages     = {252--258},
  month     = oct,
  doi       = {10.1007/s00221-004-1899-9},
  pmid      = {15112119}
}

@article{Drugowitsch2014,
  author = {Drugowitsch, Jan and DeAngelis, Gregory C. and
            Klier, Eliana M. and Angelaki, Dora E. and
            Pouget, Alexandre},
  title = {Optimal multisensory decision-making in a reaction-time task},
  journal = {eLife},
  year = {2014},
  volume = {3},
  pages = {e03005},
  doi = {10.7554/eLife.03005}
}

@article{brady2025modelling,
	author = {Brady, Rebecca M. and Butler, John S.},
	title = {Modelling Audio-Visual Reaction Time with Recurrent Mean-Field Networks},
	elocation-id = {2025.05.29.656149},
	year = {2025},
	doi = {10.1101/2025.05.29.656149},
	publisher = {Cold Spring Harbor Laboratory},
	URL = {https://www.biorxiv.org/content/early/2025/06/01/2025.05.29.656149},
	eprint = {https://www.biorxiv.org/content/early/2025/06/01/2025.05.29.656149.full.pdf},
	journal = {bioRxiv}
}

@article{butler2015optimal,
  title={Optimal visual--vestibular integration under conditions of conflicting intersensory motion profiles},
  author={Butler, John S and Campos, Jennifer L and B{\"u}lthoff, Heinrich H},
  journal={Experimental brain research},
  volume={233},
  number={2},
  pages={587--597},
  year={2015},
  publisher={Springer}
}

@article{butler2011common,
  title={Common or redundant neural circuits for duration processing across audition and touch},
  author={Butler, John S and Molholm, Sophie and Fiebelkorn, Ian C and Mercier, Manuel R and Schwartz, Theodore H and Foxe, John J},
  journal={The Journal of Neuroscience},
  volume={31},
  number={9},
  pages={3400--3406},
  year={2011},
  publisher={Society for Neuroscience}
}

@article{yau2010separate,
  title={Separate mechanisms for audio-tactile pitch and loudness interactions},
  author={Yau, Jeffrey M and Weber, Alison I and Bensmaia, Sliman J},
  journal={Frontiers in psychology},
  volume={1},
  pages={160},
  year={2010},
  publisher={Frontiers Research Foundation}
}

@article{yau2009temporal,
  title={Temporal frequency channels are linked across audition and touch},
  author={Yau, Jeffrey M and Olenczak, Jonathon B and Dammann, John F and Bensmaia, Sliman J},
  journal={Current biology},
  volume={19},
  number={7},
  pages={561--566},
  year={2009},
  publisher={Elsevier}
}

@article{FROUDISTWALSH2026,
title = {Mechanistic Computational Models of Cognition Across Scales and Species},
journal = {Biological Psychiatry},
year = {2026},
issn = {0006-3223},
doi = {https://doi.org/10.1016/j.biopsych.2026.08.017},
url = {https://www.sciencedirect.com/science/article/pii/S0006322326015131},
author = {Sean Froudist-Walsh and Tsvetoslav G. Ivanov and Teo Fantacci and Loic Magrou and Harper Cho and Eleonora Russo and Alexandra N. Busch and Lyle E. Muller and John Hongyu Meng and John D. Murray and Xiao-Jing Wang}
}

@article{Anastasio2000BayesSC,
  author    = {Anastasio, Thomas J. and Patton, Paul E. and Belkacem-Boussaid, Kamel},
  title     = {Using Bayes' Rule to Model Multisensory Enhancement in the Superior Colliculus},
  journal   = {Neural Computation},
  year      = {2000},
  volume     = {12},
  number     = {5},
  pages      = {1165--1187},
  month      = may,
  doi        = {10.1162/089976600300015547},
  pmid       = {10905812}
}

@article{diederich2004bimodal,
  title={Bimodal and trimodal multisensory enhancement: effects of stimulus onset and intensity on reaction time},
  author={Diederich, Adele and Colonius, Hans},
  journal={Perception \& psychophysics},
  volume={66},
  number={8},
  pages={1388--1404},
  year={2004},
  publisher={Springer}
}

@article{SHAW2020122,
title = {Operating in a Multisensory Context: Assessing the Interplay Between Multisensory Reaction Time Facilitation and Inter-sensory Task-switching Effects},
journal = {Neuroscience},
volume = {436},
pages = {122-135},
year = {2020},
issn = {0306-4522},
doi = {https://doi.org/10.1016/j.neuroscience.2020.04.013},
url = {https://www.sciencedirect.com/science/article/pii/S0306452220302359},
author = {Luke H. Shaw and Edward G. Freedman and Michael J. Crosse and Eric Nicholas and Allen M. Chen and Matthew S. Braiman and Sophie Molholm and John J. Foxe}
}

@article{alais2004ventriloquist,
  title={The ventriloquist effect results from near-optimal bimodal integration},
  author={Alais, David and Burr, David},
  journal={Current biology},
  volume={14},
  number={3},
  pages={257--262},
  year={2004},
  publisher={Elsevier}
}

@article{butler2010bayesian,
  title={Bayesian integration of visual and vestibular signals for heading},
  author={Butler, John S and Smith, Stuart T and Campos, Jennifer L and B{\"u}lthoff, Heinrich H},
  journal={Journal of vision},
  volume={10},
  number={11},
  pages={23--23},
  year={2010},
  publisher={The Association for Research in Vision and Ophthalmology}
}

@article{butler2011role,
  title={The role of stereo vision in visual--vestibular integration},
  author={Butler, John S and Campos, Jennifer L and B{\"u}lthoff, Heinrich H and Smith, Stuart T},
  journal={Seeing and perceiving},
  volume={24},
  number={5},
  pages={453--470},
  year={2011},
  publisher={Brill}
}

@article{roberts2024semantic,
  title={Semantic congruency modulates the speed-up of multisensory responses},
  author={Roberts, Kalvin and Jentzsch, Ines and Otto, Thomas U},
  journal={Scientific Reports},
  volume={14},
  number={1},
  pages={567},
  year={2024},
  publisher={Nature Publishing Group UK London}
}

@article{otto2013principles,
  title={Principles of multisensory behavior},
  author={Otto, Thomas U and Dassy, Brice and Mamassian, Pascal},
  journal={Journal of Neuroscience},
  volume={33},
  number={17},
  pages={7463--7474},
  year={2013},
  publisher={Society for Neuroscience}
}

@article{colonius2017measuring,
  title={Measuring multisensory integration: from reaction times to spike counts},
  author={Colonius, Hans and Diederich, Adele},
  journal={Scientific reports},
  volume={7},
  number={1},
  pages={3023},
  year={2017},
  publisher={Nature Publishing Group UK London}
}

@article{jones2016,
  author = {Jones, Pete R},
  journal = {Journal of Mathematical Psychology},
  year = {2016},
  pages = {117--139},
  title = {A tutorial on cue combination and {Signal} {Detection} {Theory}: Using changes in sensitivity to evaluate how observers integrate sensory information},
  volume = {73},
}

@article{stein2008multisensory,
  title={Multisensory integration: current issues from the perspective of the single neuron},
  author={Stein, Barry E and Stanford, Terrence R},
  journal={Nature reviews neuroscience},
  volume={9},
  number={4},
  pages={255--266},
  year={2008},
  publisher={Nature Publishing Group UK London}
}

@article{brandwein2013development,
  title={The development of multisensory integration in high-functioning autism: high-density electrical mapping and psychophysical measures reveal impairments in the processing of audiovisual inputs},
  author={Brandwein, Alice B and Foxe, John J and Butler, John S and Russo, Natalie N and Altschuler, Ted S and Gomes, Hilary and Molholm, Sophie},
  journal={Cerebral cortex},
  volume={23},
  number={6},
  pages={1329--1341},
  year={2013},
  publisher={Oxford Academic}
}

@article{magosso2008theoretical,
  title={A theoretical study of multisensory integration in the superior colliculus by a neural network model},
  author={Magosso, Elisa and Cuppini, Cristiano and Serino, Andrea and Di Pellegrino, Giuseppe and Ursino, Mauro},
  journal={Neural Networks},
  volume={21},
  number={6},
  pages={817--829},
  year={2008},
  month={Aug},
  doi={10.1016/j.neunet.2008.06.003},
  pmid={18657393},
  note={Epub 2008 Jun 22},
  publisher={Elsevier}
}

@article{terada2024chaotic,
  title={Chaotic neural dynamics facilitate probabilistic computations through sampling},
  author={Terada, Yuki and Toyoizumi, Taro},
  journal={Proceedings of the National Academy of Sciences},
  volume={121},
  number={18},
  pages={e2312992121},
  year={2024},
  publisher={National Academy of Sciences}
}

@article{sato2007bayesian,
  title={Bayesian inference explains perception of unity and ventriloquism aftereffect: identification of common sources of audiovisual stimuli},
  author={Sato, Yasuhiro and Toyoizumi, Taro and Aihara, Kazuyuki},
  journal={Neural Computation},
  volume={19},
  number={12},
  pages={3335--3355},
  year={2007},
  publisher={MIT Press}
}

@article{beck2011,
  title={Marginalization in neural circuits with divisive normalization},
  author={Beck, Jeffrey M and Latham, Peter E and Pouget, Alexandre},
  journal={Journal of Neuroscience},
  volume={31},
  number={43},
  pages={15310--15319},
  year={2011},
  publisher={Society for Neuroscience}
}

@article{Ohshiro2011,
  author = {Ohshiro, Tomokazu and Angelaki, Dora E and DeAngelis, Gregory C},
  journal = {Nature Neuroscience},
  number = {6},
  year = {2011},
  month = {may 8},
  pages = {775--782},
  title = {A normalization model of multisensory integration},
  volume = {14},
  doi = {10.1038/nn.2815},
  issn = {1097-6256},
  url = {http://dx.doi.org/10.1038/nn.2815},
}

@article{Kording2007,
  author = {K{\" o}rding, Konrad P. and Beierholm, Ulrik and Ma, Wei Ji and Quartz, Steven and Tenenbaum, Joshua B. and Shams, Ladan},
  journal = {PLoS ONE},
  number = {9},
  year = {2007},
  month = {sep 26},
  pages = {e943},
  title = {Causal {Inference} in {Multisensory} {Perception}},
  volume = {2},
  doi = {10.1371/journal.pone.0000943},
  issn = {1932-6203},
  url = {http://dx.doi.org/10.1371/journal.pone.0000943},
}

@article{Parise2016,
  author = {Parise, Cesare V. and Ernst, Marc O.},
  journal = {Nature Communications},
  number = {1},
  year = {2016},
  month = {jun 6},
  title = {Correlation detection as a general mechanism for multisensory integration},
  volume = {7},
  doi = {10.1038/ncomms11543},
  issn = {2041-1723},
  url = {http://dx.doi.org/10.1038/ncomms11543},
}

@article{pesnot2022,
  title={Multisensory correlation computations in the human brain identified by a time-resolved encoding model},
  author={Pesnot Lerousseau, Jacques and Parise, Cesare V and Ernst, Marc O and van Wassenhove, Virginie},
  journal={Nature communications},
  volume={13},
  number={1},
  pages={2489},
  year={2022},
  publisher={Nature Publishing Group UK London}
}

@article{parise2025,
  title={Multisensory integration operates on correlated input from unimodal transient channels},
  author={Parise, Cesare V and Ernst, Marc O},
  journal={eLife},
  volume={12},
  pages={RP90841},
  year={2025},
  publisher={eLife Sciences Publications Limited}
}

@article{ernst2002,
  title={Humans integrate visual and haptic information in a statistically optimal fashion},
  author={Ernst, Marc O and Banks, Martin S},
  journal={Nature},
  volume={415},
  number={6870},
  pages={429--433},
  year={2002},
  publisher={Nature Publishing Group UK London}
}

@article{Ghosh2024,
  title={Nonlinear fusion is optimal for a wide class of multisensory tasks},
  author={Ghosh, Marcus and B{\'e}na, Gabriel and Bormuth, Volker and Goodman, Dan FM},
  journal={PLOS Computational Biology},
  volume={20},
  number={7},
  pages={e1012246},
  year={2024},
  publisher={Public Library of Science San Francisco, CA USA}
}

@article{anil2025,
  title={Fusing multisensory signals across channels and time},
  author={Anil, Swathi and Goodman, Dan FM and Ghosh, Marcus},
  journal={PLOS Computational Biology},
  volume={21},
  number={6},
  pages={e1013125},
  year={2025},
  publisher={Public Library of Science San Francisco, CA USA}
}

@article{ghosh2026,
  title={Partial recurrence can enable robust and efficient computation},
  author={Ghosh, Marcus and Goodman, Dan FM},
  journal={Communications AI \& Computing},
  volume={1},
  number={1},
  pages={11},
  year={2026},
  publisher={Nature Publishing Group UK, London}
}

@article{ghosh2025,
  title={Not playing around: Why neuroscience needs toy models},
  author={Ghosh, Marcus},
  journal={The Transmitter},
  year={2025},
}

@article{Nagrani2021,
  title={Attention bottlenecks for multimodal fusion},
  author={Nagrani, Arsha and Yang, Shan and Arnab, Anurag and Jansen, Aren and Schmid, Cordelia and Sun, Chen},
  journal={Advances in neural information processing systems},
  volume={34},
  pages={14200--14213},
  year={2021}
}

@article{rideaux2021,
  title={How multisensory neurons solve causal inference},
  author={Rideaux, Reuben and Storrs, Katherine R and Maiello, Guido and Welchman, Andrew E},
  journal={Proceedings of the National Academy of Sciences},
  volume={118},
  number={32},
  pages={e2106235118},
  year={2021},
  publisher={National Acad Sciences}
}

@article{gu2008neural,
  title={Neural correlates of multisensory cue integration in macaque MSTd},
  author={Gu, Yong and Angelaki, Dora E and DeAngelis, Gregory C},
  journal={Nature neuroscience},
  volume={11},
  number={10},
  pages={1201--1210},
  year={2008},
  publisher={Nature Publishing Group US New York}
}

@article{farahmandi2024,
  title={Multisensory integration across reference frames with additive feed-forward networks},
  author={Farahmandi, Arefeh and Abedi Khoozani, Parisa and Blohm, Gunnar},
  journal={bioRxiv},
  pages={2024--11},
  year={2024},
  publisher={Cold Spring Harbor Laboratory}
}

@article{tugsbayar2024,
  title={Top-down feedback matters: Functional impact of brainlike connectivity motifs on audiovisual integration},
  author={Tugsbayar, Mashbayar and Li, Mingze and Muller, Eilif B and Richards, Blake},
  journal={bioRxiv},
  pages={2024--10},
  year={2024},
  publisher={Cold Spring Harbor Laboratory}
}

@article{eckstein2026,
  title={Hybrid neural--cognitive models reveal how memory shapes human reward learning},
  author={Eckstein, Maria K and Summerfield, Christopher and Daw, Nathaniel D and Miller, Kevin J},
  journal={Nature Human Behaviour},
  pages={1--16},
  year={2026},
  publisher={Nature Publishing Group UK London}
}

@article{castro2025,
  title={Discovering Symbolic Cognitive Models from Human and Animal Behavior},
  author={Castro, Pablo Samuel and Tomasev, Nenad and Anand, Ankit and Sharma, Navodita and Mohanta, Rishika and Dev, Aparna and Perlin, Kuba and Jain, Siddhant and Levin, Kyle and {\'E}ltet{\H{o}}, No{\'e}mi and others},
  journal={bioRxiv},
  pages={2025--02},
  year={2025},
  publisher={Cold Spring Harbor Laboratory}
}

@article{oess2020,
  title={From near-optimal bayesian integration to neuromorphic hardware: A neural network model of multisensory integration},
  author={Oess, Timo and L{\"o}hr, Maximilian PR and Schmid, Daniel and Ernst, Marc O and Neumann, Heiko},
  journal={Frontiers in Neurorobotics},
  volume={14},
  pages={29},
  year={2020},
  publisher={Frontiers Media SA}
}

@article{miller1982,
  title={Divided attention: Evidence for coactivation with redundant signals},
  author={Miller, Jeff},
  journal={Cognitive psychology},
  volume={14},
  number={2},
  pages={247--279},
  year={1982},
  publisher={Elsevier}
}

@article{raab1962,
  title={Statistical facilitation of simple reaction times.},
  author={Raab, David H},
  journal={Transactions of the New York Academy of Sciences},
  year={1962}
}

@article{otto2017,
  title={Multisensory decisions: the test of a race model, its logic, and power},
  author={Otto, Thomas U and Mamassian, Pascal},
  journal={Multisensory Research},
  volume={30},
  number={1},
  pages={1--24},
  year={2017},
  publisher={Brill}
}

@article{roberts2025,
  title={When one race is not enough: a relay model explains multisensory response times},
  author={Roberts, Kalvin and Otto, Thomas U},
  journal={bioRxiv},
  pages={2025--07},
  year={2025},
  publisher={Cold Spring Harbor Laboratory}
}

@article{ursino2009,
  title={Multisensory integration in the superior colliculus: a neural network model},
  author={Ursino, Mauro and Cuppini, Cristiano and Magosso, Elisa and Serino, Andrea and Di Pellegrino, Giuseppe},
  journal={Journal of computational neuroscience},
  volume={26},
  number={1},
  pages={55--73},
  year={2009},
  publisher={Springer}
}

@article{meredith1986visual,
  title={Visual, auditory, and somatosensory convergence on cells in superior colliculus results in multisensory integration},
  author={Meredith, M Alex and Stein, Barry E},
  journal={Journal of neurophysiology},
  volume={56},
  number={3},
  pages={640--662},
  year={1986},
  publisher={American Physiological Society Bethesda, MD}
}

@article{macneilage2007bayesian,
  title={A Bayesian model of the disambiguation of gravitoinertial force by visual cues},
  author={MacNeilage, Paul R and Banks, Martin S and Berger, Daniel R and B{\"u}lthoff, Heinrich H},
  journal={Experimental Brain Research},
  volume={179},
  number={2},
  pages={263--290},
  year={2007},
  publisher={Springer}
}

@incollection{yuille1993bayesian,
    author={Yuille, Alan L and B{\"u}lthoff, Heinrich H} ,
    title={Bayesian decision theory and psychophysics},
    booktitle={Perception as Bayesian inference},
    publisher={Cambridge University Press},
    year={1993} 
}

@article{paszke2019,
  title={Pytorch: An imperative style, high-performance deep learning library},
  author={Paszke, Adam and Gross, Sam and Massa, Francisco and Lerer, Adam and Bradbury, James and Chanan, Gregory and Killeen, Trevor and Lin, Zeming and Gimelshein, Natalia and Antiga, Luca and others},
  journal={Advances in neural information processing systems},
  volume={32},
  year={2019}
}

@article{liang2022,
  title={Foundations and recent trends in multimodal machine learning: Principles, challenges, and open questions},
  author={Liang, Paul Pu and Zadeh, Amir and Morency, Louis-Philippe},
  journal={arXiv preprint arXiv:2209.03430},
  volume={4},
  number={5},
  year={2022}
}

@article{coen2023,
  title={Mouse frontal cortex mediates additive multisensory decisions},
  author={Coen, Philip and Sit, Timothy PH and Wells, Miles J and Carandini, Matteo and Harris, Kenneth D},
  journal={Neuron},
  volume={111},
  number={15},
  pages={2432--2447},
  year={2023},
  publisher={Elsevier}
}

@book{bishop2024,
author = {Christopher M. Bishop and Hugh Bishop},
title = {Deep Learning: Foundations and Concepts},
year = {2024},
publisher = {Springer}
}

@article{jonas2017,
  title={Could a neuroscientist understand a microprocessor?},
  author={Jonas, Eric and Kording, Konrad Paul},
  journal={PLoS computational biology},
  volume={13},
  number={1},
  pages={e1005268},
  year={2017},
  publisher={Public Library of Science}
}

@article{fakhar2022,
  title={Systematic perturbation of an artificial neural network: A step towards quantifying causal contributions in the brain},
  author={Fakhar, Kayson and Hilgetag, Claus C},
  journal={PLOS Computational Biology},
  volume={18},
  number={6},
  pages={e1010250},
  year={2022},
  publisher={Public Library of Science San Francisco, CA USA}
}

@article{senkowski2024,
  title={Multi-timescale neural dynamics for multisensory integration},
  author={Senkowski, Daniel and Engel, Andreas K},
  journal={Nature Reviews Neuroscience},
  volume={25},
  number={9},
  pages={625--642},
  year={2024},
  publisher={Nature Publishing Group UK London}
}

@article{talsma2015,
  title={Predictive coding and multisensory integration: an attentional account of the multisensory mind},
  author={Talsma, Durk},
  journal={Frontiers in integrative neuroscience},
  volume={9},
  pages={19},
  year={2015},
  publisher={Frontiers Media SA}
}

@article{elman1990,
  title={Finding structure in time},
  author={Elman, Jeffrey L},
  journal={Cognitive science},
  volume={14},
  number={2},
  pages={179--211},
  year={1990},
  publisher={Wiley Online Library}
}

@article{chandrasekaran2017,
  title={Computational principles and models of multisensory integration},
  author={Chandrasekaran, Chandramouli},
  journal={Current opinion in neurobiology},
  volume={43},
  pages={25--34},
  year={2017},
  publisher={Elsevier}
}

@article{bolhasani2026,
  title={Computational models of multisensory integration with recurrent neural networks: A critical review and future directions},
  author={Bolhasani, Ehsan and Aboutalebi, Seyed Hamed and Merrikhi, Yaser},
  journal={Advanced Intelligent Systems},
  volume={8},
  number={1},
  pages={2500147},
  year={2026},
  publisher={Wiley Online Library}
}

@article{yamashita2013,
  title={Recurrent network for multisensory integration-identification of common sources of audiovisual stimuli},
  author={Yamashita, Itsuki and Katahira, Kentaro and Igarashi, Yasuhiko and Okanoya, Kazuo and Okada, Masato},
  journal={Frontiers in Computational Neuroscience},
  volume={7},
  pages={101},
  year={2013},
  publisher={Frontiers Media SA}
}

@article{wierda2025,
  title={Diverse and flexible behavioral strategies arise in recurrent neural networks trained on multisensory decision making},
  author={Wierda, Thomas S and Dora, Shirin and Pennartz, Cyriel MA and Mejias, Jorge F},
  journal={PLOS Computational Biology},
  volume={21},
  number={10},
  pages={e1013559},
  year={2025},
  publisher={Public Library of Science San Francisco, CA USA}
}

@article{patterson2023,
  title={Empirical design in reinforcement learning},
  author={Patterson, Andrew and Neumann, Samuel and White, Martha and White, Adam},
  journal={arXiv preprint arXiv:2304.01315},
  year={2023}
}

@book{ma2023,
  title={Bayesian models of perception and action: An introduction},
  author={Ma, Wei Ji and Kording, Konrad Paul and Goldreich, Daniel},
  year={2023},
  publisher={MIT press}
}

@article{ji2025,
  title={Discovering cognitive strategies with tiny recurrent neural networks},
  author={Ji-An, Li and Benna, Marcus K and Mattar, Marcelo G},
  journal={Nature},
  volume={644},
  number={8078},
  pages={993--1001},
  year={2025},
  publisher={Nature Publishing Group UK London}
}

@article{bizley2016,
  title={Where are multisensory signals combined for perceptual decision-making?},
  author={Bizley, Jennifer K and Jones, Gareth P and Town, Stephen M},
  journal={Current opinion in neurobiology},
  volume={40},
  pages={31--37},
  year={2016},
  publisher={Elsevier}
}

@article{barnum2020,
  title={On the benefits of early fusion in multimodal representation learning},
  author={Barnum, George and Talukder, Sabera and Yue, Yisong},
  journal={arXiv preprint arXiv:2011.07191},
  year={2020}
}

@inproceedings{gadzicki2020,
  title={Early vs late fusion in multimodal convolutional neural networks},
  author={Gadzicki, Konrad and Khamsehashari, Razieh and Zetzsche, Christoph},
  booktitle={2020 IEEE 23rd international conference on information fusion (FUSION)},
  pages={1--6},
  year={2020},
  organization={IEEE}
}

@article{vaswani2017,
  title={Attention is all you need},
  author={Vaswani, Ashish and Shazeer, Noam and Parmar, Niki and Uszkoreit, Jakob and Jones, Llion and Gomez, Aidan N and Kaiser, {\L}ukasz and Polosukhin, Illia},
  journal={Advances in neural information processing systems},
  volume={30},
  year={2017}
}

@article{marder2011,
  title={Multiple models to capture the variability in biological neurons and networks},
  author={Marder, Eve and Taylor, Adam L},
  journal={Nature neuroscience},
  volume={14},
  number={2},
  pages={133--138},
  year={2011},
  publisher={Nature Publishing Group US New York}
}

@article{edelman2001,
  title={Degeneracy and complexity in biological systems},
  author={Edelman, Gerald M and Gally, Joseph A},
  journal={Proceedings of the national academy of sciences},
  volume={98},
  number={24},
  pages={13763--13768},
  year={2001},
  publisher={National Acad Sciences}
}

@article{peters2025,
  title={Generative adversarial collaborations: A new model of scientific discourse},
  author={Peters, Benjamin and Blohm, Gunnar and Haefner, Ralf and Isik, Leyla and Kriegeskorte, Nikolaus and Lieberman, Jennifer S and Ponce, Carlos R and Roig, Gemma and Peters, Megan AK},
  journal={Trends in Cognitive Sciences},
  volume={29},
  number={1},
  pages={1--4},
  year={2025},
  publisher={Elsevier}
}

@article{blohm2024,
  title={Generative Adversarial Collaborations: A practical guide for conference organizers and participating scientists},
  author={Blohm, Gunnar and Peters, Benjamin and Haefner, Ralf and Isik, Leyla and Kriegeskorte, Nikolaus and Lieberman, Jennifer S and Ponce, Carlos R and Roig, Gemma and Peters, Megan AK},
  journal={arXiv preprint arXiv:2402.12604},
  year={2024}
}

@book{Marr1982,
  title     = {Vision: A Computational Investigation into the Human Representation and Processing of Visual Information},
  author    = {Marr, David},
  year      = {1982},
  publisher = {W. H. Freeman},
  address   = {San Francisco, CA}
}

@article{pillow2024,
  title={Cross Talk opposing view: Marr's three levels of analysis are not useful as a framework for neuroscience},
  author={Pillow, Jonathan W},
  journal={The Journal of Physiology},
  volume={602},
  number={9},
  pages={1915--1917},
  year={2024}
}

@article{lengyel2024,
  title={Crosstalk rebuttal: Marr's three levels of analysis are not useful as a framework for neuroscience},
  author={Lengyel, M{\'a}t{\'e}},
  journal={The Journal of Physiology},
  volume={602},
  number={9},
  year={2024},
}

@article{glaser2020,
  title={Machine learning for neural decoding},
  author={Glaser, Joshua I and Benjamin, Ari S and Chowdhury, Raeed H and Perich, Matthew G and Miller, Lee E and Kording, Konrad P},
  journal={eneuro},
  volume={7},
  number={4},
  year={2020},
  publisher={Society for Neuroscience}
}

@article{richards2019,
  title={A deep learning framework for neuroscience},
  author={Richards, Blake A and Lillicrap, Timothy P and Beaudoin, Philippe and Bengio, Yoshua and Bogacz, Rafal and Christensen, Amelia and Clopath, Claudia and Costa, Rui Ponte and de Berker, Archy and Ganguli, Surya and others},
  journal={Nature neuroscience},
  volume={22},
  number={11},
  pages={1761--1770},
  year={2019},
  publisher={Nature Publishing Group US New York}
}

@article{doerig2023,
  title={The neuroconnectionist research programme},
  author={Doerig, Adrien and Sommers, Rowan P and Seeliger, Katja and Richards, Blake and Ismael, Jenann and Lindsay, Grace W and Kording, Konrad P and Konkle, Talia and Van Gerven, Marcel AJ and Kriegeskorte, Nikolaus and others},
  journal={Nature Reviews Neuroscience},
  volume={24},
  number={7},
  pages={431--450},
  year={2023},
  publisher={Nature Publishing Group UK London}
}

@article{romera2024,
  title={Mathematical discoveries from program search with large language models},
  author={Romera-Paredes, Bernardino and Barekatain, Mohammadamin and Novikov, Alexander and Balog, Matej and Kumar, M Pawan and Dupont, Emilien and Ruiz, Francisco JR and Ellenberg, Jordan S and Wang, Pengming and Fawzi, Omar and others},
  journal={Nature},
  volume={625},
  number={7995},
  pages={468--475},
  year={2024},
  publisher={Nature Publishing Group UK London}
}

@article{crilly2025,
  title={Ten simple rules for navigating AI in science},
  author={Crilly, Aidan and Malivert, Alice and J{\o}rgensen, Andreas Christ S{\o}lvsten and Heaney, Claire E and Vera Gonzalez, Gema I and Ghosh, Marcus and Fernandez Perez, Manolo and Mieskolainen, Mikael M and Azzouzi, Mohammed and Li, Zhenzhu},
  journal={PLoS computational biology},
  volume={21},
  number={7},
  pages={e1013259},
  year={2025},
  publisher={Public Library of Science San Francisco, CA USA}
}

@article{zhu2026,
  title={Analyzing multisensory integration: do’s and dont’s},
  author={Zhu, Haocheng and Beierholm, Ulrik and Shams, Ladan},
  journal={bioRxiv},
  pages={2026--04},
  year={2026},
  publisher={Cold Spring Harbor Laboratory}
}

@article{falchier2002,
  title={Anatomical evidence of multimodal integration in primate striate cortex},
  author={Falchier, Arnaud and Clavagnier, Simon and Barone, Pascal and Kennedy, Henry},
  journal={Journal of Neuroscience},
  volume={22},
  number={13},
  pages={5749--5759},
  year={2002},
  publisher={Society for Neuroscience}
}

@article{driver2008,
  title={Multisensory interplay reveals crossmodal influences on ‘sensory-specific’brain regions, neural responses, and judgments},
  author={Driver, Jon and Noesselt, Toemme},
  journal={Neuron},
  volume={57},
  number={1},
  pages={11--23},
  year={2008},
  publisher={Elsevier}
}

@article{ghazanfar2006,
  title={Is neocortex essentially multisensory?},
  author={Ghazanfar, Asif A and Schroeder, Charles E},
  journal={Trends in cognitive sciences},
  volume={10},
  number={6},
  pages={278--285},
  year={2006},
  publisher={Elsevier}
}

@article{cappe2005,
  title={Heteromodal connections supporting multisensory integration at low levels of cortical processing in the monkey},
  author={Cappe, C{\'e}line and Barone, Pascal},
  journal={European Journal of Neuroscience},
  volume={22},
  number={11},
  pages={2886--2902},
  year={2005},
  publisher={Wiley Online Library}
}

@article{baltruvsaitis2018,
  title={Multimodal machine learning: A survey and taxonomy},
  author={Baltru{\v{s}}aitis, Tadas and Ahuja, Chaitanya and Morency, Louis-Philippe},
  journal={IEEE transactions on pattern analysis and machine intelligence},
  volume={41},
  number={2},
  pages={423--443},
  year={2018},
  publisher={IEEE}
}

@article{dionne2015,
  title={Multisensory integration and child neurodevelopment},
  author={Dionne-Dostie, Emmanuelle and Paquette, Natacha and Lassonde, Maryse and Gallagher, Anne},
  journal={Brain sciences},
  volume={5},
  number={1},
  pages={32--57},
  year={2015},
  publisher={MDPI}
}

@article{hornix2019,
  title={Multisensory cortical processing and dysfunction across the neuropsychiatric spectrum},
  author={Hornix, Betty E and Havekes, Robbert and Kas, Martien JH},
  journal={Neuroscience \& Biobehavioral Reviews},
  volume={97},
  pages={138--151},
  year={2019},
  publisher={Elsevier}
}

@article{HassensteinReichardt1956,
  author = {Hassenstein, Bernhard and Reichardt, Werner},
  title = {Systemtheoretische Analyse der Zeit-, Reihenfolgen- und Vorzeichenauswertung bei der Bewegungsperzeption des R{\"u}sselk{\"a}fers Chlorophanus},
  journal = {Zeitschrift f{\"u}r Naturforschung B},
  year = {1956},
  volume = {11},
  number = {9--10},
  pages = {513--524},
}

@article{beer2024,
  title={Milking a spherical cow: Toy models in neuroscience},
  author={Beer, Randall D and Barwich, Ann-Sophie and Severino, Gabriel J},
  journal={European Journal of Neuroscience},
  volume={60},
  number={10},
  pages={6359--6374},
  year={2024},
  publisher={Wiley Online Library}
}
\section*{Acknowledgements}
MG is supported by Schmidt Sciences LLC and an Imperial College Research Fellowship.

\end{document}